\documentclass[manuscript]{emulateapj}
\usepackage{amssymb}
\usepackage[export]{adjustbox}
\usepackage{amsmath}
\usepackage{booktabs}
\usepackage{hyperref}
\usepackage{multirow}
\usepackage{times}
\PassOptionsToPackage{hyphens}{url}\usepackage{hyperref}
\usepackage[hyphenbreaks]{breakurl}

\newcommand{\mcg}{MCG$-$03$-$58$-$007}
\newcommand{\degg}{\hbox{$^\circ$}}
\newcommand{\zw}{I\,Zw\,1}
\newcommand{\pg}{PG\,1448+273}
\newcommand{\xmm}{{\it XMM-Newton}}

\newcommand{\arcs}{\hbox{$^{\prime\prime}$}}

\newcommand{\ls}
{\mathrel{\hbox{\rlap{\hbox{\lower4pt\hbox{$\sim$}}}\hbox{$<$}}}}
\newcommand{\gs}
{\mathrel{\hbox{\rlap{\hbox{\lower4pt\hbox{$\sim$}}}\hbox{$>$}}}}

\received{}
\begin{document}
\title{Rapid Variability of the accretion disk wind in the narrow line Seyfert 1, PG\,1448+273.}
\shorttitle{A Variable Disk Wind in PG\,1448+273.}
\shortauthors{Reeves et al.}
\author{J. N. Reeves\altaffilmark{1,2}, V. Braito\altaffilmark{1,2,3}, D. Porquet\altaffilmark{4}, M. Laurenti\altaffilmark{5,6}, A. Lobban\altaffilmark{7}, G. Matzeu\altaffilmark{8}}
\altaffiltext{1}{Department of Physics, Institute for Astrophysics and Computational Sciences, The Catholic University of America, Washington, DC 20064, USA, email reevesjn@cua.edu}
\altaffiltext{2}{INAF, Osservatorio Astronomico di Brera, Via Bianchi 46 I-23807 Merate (LC), Italy}
\altaffiltext{3}{Dipartimento di Fisica, Universit\`a di Trento, Via Sommarive 14, Trento 38123, Italy.}
\altaffiltext{4}{Aix-Marseille Univ., CNRS, CNES, LAM, Marseille, France}
\altaffiltext{5}{Dipartimento di Fisica, Universit\`a di Roma “Tor Vergata”, Via della Ricerca Scientifica 1, 00133 Roma, Italy}
\altaffiltext{6}{Space Science Data Center, SSDC, ASI, Via del Politecnico snc, 00133 Roma, Italy}
\altaffiltext{7}{European Space Astronomy Centre (ESA/ESAC), E-28691 Villanueva de la Canada, Madrid, Spain}
\altaffiltext{8}{Department of Physics and Astronomy (DIFA), University of Bologna, Via Gobetti, 93/2, I-40129 Bologna, Italy}


\begin{abstract}

\pg\ is a luminous, nearby ($z=0.0645$), narrow line Seyfert 1 galaxy, which likely accretes close to the Eddington limit. 
\xmm\ observations of \pg\ in 2017 revealed the presence of an ultra fast outflow, 
as seen through its blueshifted iron K absorption profile, 
with an outflow velocity of about $0.1c$. Here, the first {\it NuSTAR} observation of \pg, performed in 2022 and coordinated with \xmm\ is presented, 
which shows remarkable variability of its ultra fast outflow. The average count rate is a factor of 2 lower during the last 60\,ks of the {\it NuSTAR} 
observation, where a much faster component of the ultra fast outflow was detected with a terminal velocity of $0.26\pm0.04c$. This is significantly faster 
than the outflow component which was initially detected in 2017, when overall \pg\ was observed at a lower X-ray flux and which implies 
an order of magnitude increase in the wind kinetic power between the 2017 and 2022 epochs. Furthermore, the rapid variability of the ultra fast outflow 
in 2022, on timescales down to 10\,ks, suggests we are viewing through a highly inhomogeneous disk wind in \pg, where the passage of a denser wind clump 
could account for the increase in obscuration in the last 60\,ks of the {\it NuSTAR} observation.
\end{abstract}

\keywords{galaxies: active --- quasars: individual (PG\,1448+273) --- X-rays: galaxies --- black hole physics}









\section{Introduction}

Ultra fast outflows (or UFOs) were first detected through observations of blueshifted iron K-shell absorption profiles, as observed in the 
X-ray spectra of Active Galactic Nuclei (AGN). The first known examples of these fast outflows were discovered in the 
luminous quasars, APM\,08279+5255 (\citealt{Chartas02}), PG\,1211+143 (\citealt{Pounds03}) and PDS\,456 (\citealt{Reeves03}). 
Since their initial discovery, a number of high column density ($N_{\rm H}\sim 10^{23}$\,cm$^{-2}$), ultra fast ($\sim0.1c$)
outflows have been found in luminous nearby AGN (\citealt{Tombesi10,Gofford13}). 
These fast winds span a wide velocity range of up to $\sim0.3c$, 
as seen in both PDS\,456 (\citealt{Matzeu17}) and the broad absorption line quasar, APM\,08279+5255 (\citealt{SaezChartas11}).

The mechanical power of these winds can reach a significant fraction of the Eddington limit, which may be more than sufficient to provide the mechanical feedback required by models of black hole and host galaxy co-evolution (\citealt{SilkRees98, Fabian99,DiMatteo05,King03,HopkinsElvis10}). Such black hole winds may play a crucial 
part in the regulating the growth of super massive black holes and the bulges of their host galaxies in luminous QSOs 
(\citealt{FerrareseMerritt00,Gebhardt00,Tremaine02}).

The subject of this paper is the nearby ($z=0.0645$) Narrow Line Seyfert 1, \pg. 
It is also classed as a radio-quiet QSO \citep{SchmidtGreen83} and has narrow permitted lines, e.g. 
H$\beta$ FWHM of 1330\,km\,s$^{-1}$ \citep{Grupe04}.  
Its bolometric luminosity is estimated to be $L_{\rm bol}=2-3\times10^{45}$\,erg\,s$^{-1}$, while its black hole mass lies in the range from
 $M_{\rm BH}=0.9-2.0\times10^{7}$\,M$_{\odot}$ \citep{VestergaardPeterson06, Shen11}. 
 Recently, from a survey of SDSS quasars, \citet{Rakshit20} estimate a black hole mass for \pg\ (based upon a single epoch H$\beta$ measurement) of $\log(M/M_{\odot})=7.14$ and a
 bolometric luminosity of $\log (L/({\rm erg\,s}^{-1}))=45.24$. 
 These measurements imply that \pg\ is likely to accrete near the Eddington limit. 
 Such high accretion rate AGN are prime candidates for driving a fast disk wind. 
Indeed, there are now several examples of fast winds in other NLS1s, reaching velocities 
 of up to $0.3c$; e.g. IRAS\,13224$-$3809 \citep{Parker17}, PG\,1211+143 \citep{Pounds03}, 
 IRAS\,17020+4544 \citep{Longinotti15}, 1H\,0707$-$495 \citep{Kosec18}, \zw\ \citep{ReevesBraito19}. 
Thus as the AGN reach the Eddington limit, they may become more efficient at powering an accretion disk wind. 

NLS1s are highly variable in X-rays \citep{Boller96} and \pg\ is no exception to this, showing both order of magnitude and short timescale variability (e.g. see Figure~4, \citealt{Laurenti21}). Variability can provide vital insights into the nature of AGN disk winds and their physical characteristics. The wind velocity can react to the X-ray flux, becoming faster vs luminosity, potentially as a result 
of radiation driving (e.g. PDS 456; \citealt{Matzeu17}). The wind ionization can also increase with flux, decreasing the wind opacity  \citep{Pinto18}. Finally, increases in wind obscuration can occur, as a result of ejecta passing across the sightline; e.g. PG\,1211+143; \citep{ReevesLobbanPounds18}, \mcg\ \citep{Braito18}. 
  
A 2017 \xmm\ observation (75\,ks net pn exposure) has first established the presence of a fast wind in \pg\ \citep{Kosec20, Laurenti21}, measured by a broad and blue-shifted iron K absorption trough observed at 7.5\,keV; e.g. see Figure~1, \citealt{Laurenti21} and Figure~3, \citealt{Kosec20}. 
If the absorption profile is associated with the strong $1s\rightarrow2p$ resonance lines of  He and H-like iron, then the implied outflow velocity is about $0.1c$. 
The equivalent width of the absorption profile is also high, with ${\rm EW}=-410\pm80$\,eV \citep{Laurenti21}, one of the highest found for UFOs to date (c.f. \citealt{Tombesi10, Gofford13}) 
and implies that the column density may reach up to $N_{\rm H}=10^{24}$\,cm$^{-2}$. 
As shown by \citet{Laurenti21}, the iron line profile in the 2017 observation can be fitted with a disk wind profile, via the WINE code of \citet{Luminari18}, which is reminiscent of the wind profile observed in the high luminosity QSO PDS\,456 \citep{Nardini15}. Overall the X-ray flux in 2017 was low ($F_{2-10\,{\rm keV}}=1.3\times10^{-12}$\,erg\,cm$^{-2}$\,s$^{-1}$), compared to the {\it Swift} monitoring observations, where the \xmm\ observation occurred just prior to a pronounced dip in the Swift lightcurve \citep{Laurenti21}. 

In this paper we present the first {\it NuSTAR} observation of \pg\ (250\,ks duration, 130\,ks net exposure, Table~\ref{tab:obs}) which occurred in January 2022 and revealed strong X-ray variability.  
The first part of the observation (Slice A) is seen at a much higher flux ($F_{2-10\,{\rm keV}}=4.8\times10^{-12}$\,erg\,cm$^{-2}$\,s$^{-1}$), with a generally featureless X-ray spectrum and is coincident with a 70\,ks \xmm\ exposure. However the last 60\,ks of the {\it NuSTAR} observation (Slice B) 
coincided with a rapid decline in flux and the emergence of a deep, iron K absorption profile at 9\,keV in the AGN rest-frame. 

The motivation of this paper is to explore in detail the wind variability of \pg\ both on short timescales within the 2022 campaign and on long timescales, via the comparison with the earlier 2017 observation. 
The paper is organized as follows. 
In Section~2, the observations and data reduction are described, 
while in Section~3 the variability is quantified within the long {\it NuSTAR} observation, including the energy dependent variability in the form of the fraction rms (or $F_{\rm var}$) spectra \citep{Vaughan03, Parker20, Igo20}. 
The spectra are fitted with simple \textsc{xstar} models in Section~4, above 3 keV in order to compare the properties of the iron K line profile,  
including the comparison between the 2017 and 2022 epochs. In Section~5 the full-band (0.3--30\,keV) multi-epoch X-ray spectra are modeled, 
with the radiative transfer disk wind model of \citet{Sim08,Sim10a,Sim10b} and utilizing the subsequent tables of wind models computed by \citet{Matzeu22}. In Section~6, the \xmm\ soft X-ray RGS grating spectra are compared between the 2017 and 2022 epochs, in order to understand the wind properties
at higher resolution in the soft X-ray band. As will be shown in Section 6, the former reveals putative blue-shifted absorption troughs from the disk wind \citep{Kosec20}, while the brighter 2022 
spectrum reveals that the wind opacity decreases with increasing flux. The possible origins of the wind variability, on both short and long timescales, are then discussed in Section~7.  
Throughout the paper, 90\% confidence intervals for 1 parameter of interested are adopted for the uncertainties (or $\Delta \chi^2=2.7$), while parameters are stated in the AGN rest frame 
at $z=0.0645$. The standard $\Lambda$CDM cosmology ($H_0 = 70$\,km\,s${-1}$\, Mpc\,$^{-1}$, $\Omega_m = 0.3$, $\Omega_{\Lambda} = 0.7$) is adopted throughout the paper.

\section{Observations and Data Reduction}

\pg\ was observed with {\it NuSTAR} \citep{Harrison13} in January 2022 for a total duration of 253\,ks and a net exposure per FPMA/B module of 130\,ks. 
Part of the {\it NuSTAR} observation was performed simultaneously with \xmm, where with respect to the EPIC-pn detector, the total observation duration was 74\,ks. 
The details of the 2022 {\it XMM-Newton} and {\it NuSTAR} observations are listed in Table~\ref{tab:obs}.
The EPIC-MOS \citep{Turner01} exposures were performed in Small Window mode and the EPIC-pn \citep{Struder01} exposure in Large window mode in order to minimize pile-up. 
The 2017 \xmm\ observation was also reanalyzed, which occurred over a single satellite orbit over a duration of 120\,ks and in Small Window Mode, but which was not performed simultaneously 
with {\it NuSTAR}. 
The exposures are also listed in Table~\ref{tab:obs}.

\begin{deluxetable}{lccc}
\tablecaption{Observation log of \pg.}
\tablewidth{250pt}
\tablehead{
\colhead{Instrument} & \colhead{Start Date (UT)} & \colhead{Exp (ks)$^a$} & \colhead{Rate(s$^{-1}$)$^b$} }
\startdata
FPMA/B & 2022/01/04 10:51:09 & 130.0 & $0.130\pm0.001$ \\
\hline\\
XMM/pn & 2022/01/05 06:03:45 & 59.3 & $9.462\pm0.013$ \\
XMM/MOS & -- & 69.9 &  $1.938\pm0.004$ \\
XMM/RGS & -- & 75.5 &  $0.336\pm0.002$ \\
\hline\\
XMM/pn & 2017/01/24 06:15:46 & 76.2 & $2.051\pm0.005$ \\
XMM/MOS & -- & 108.6 &  $0.425\pm0.002$ \\
XMM/RGS & -- & 115.8 &  $0.068\pm0.001$ 
\enddata
\tablenotetext{a}{Net exposure, correcting for background screening and detector deadtime.}
\tablenotetext{b}{Average net count rates per MOS, RGS or {\it NuSTAR} FPM module.}
\label{tab:obs}
\end{deluxetable}

The observations were processed using the \textsc{nustardas} v2.1.2, 
{\it XMM-Newton} \textsc{sas} v20.0 and \textsc{heasoft} v6.30 software. 
{\it NuSTAR} source spectra were extracted using a 52\arcs\ circular region centered on the source and two background  circular regions with  a 52\arcs\  radius and clear from stray light. 
{\it XMM-Newton} EPIC-pn spectra were extracted from single and double events, using a 36\arcs\ source region and $2\times36\arcs$\ background regions on the same chip. 
EPIC-MOS spectra were extracted using patterns 0--12, using a 30\arcs\ source region and $2\times36\arcs$\ background regions. The spectra and responses from the individual FPMA and FPMB detectors on-board {\it NuSTAR} 
were combined into a single spectrum after they were first checked for consistency and yielded a net source count rate of  $0.130\pm0.001$\,cts\,s$^{-1}$ per detector. The {\it NuSTAR} spectra were utilized over the 3--30\,keV band; above 30\,keV the source spectrum becomes background dominated as the source count rate declines. The total background rate over this band is 6\% of the source rate. 
All the spectra are binned to at least 100 counts per bin to enable the use of $\chi^2$ statistics. 

After background subtraction, the 2022 EPIC-pn spectrum resulted in a mean net count rate of $9.462\pm0.008$\,cts\,s$^{-1}$ over the full 0.3--10 keV band (or $0.259\pm0.002$\,cts\,s$^{-1}$ from 3--10\,keV) and a 
net exposure of 59.3\,ks after correcting for detector deadtime.  The background level was very low, $<0.4$\% of the net source rate and significant background flaring only occurred during the first $\sim5$\,ks 
of the observation, which were filtered out from the final spectrum.  
In contrast, the 2017 \xmm\ observation 
has lower overall count rates (see Table~\ref{tab:obs}), with a net count rate of $2.051\pm0.005$\,cts\,s$^{-1}$ for the EPIC-pn detector over the 0.3--10\,keV band ($0.093\pm0.001$\,cts\,s$^{-1}$ from 3--10\,keV).  

\begin{figure}
\begin{center}
\rotatebox{-90}{\includegraphics[height=8.5cm]{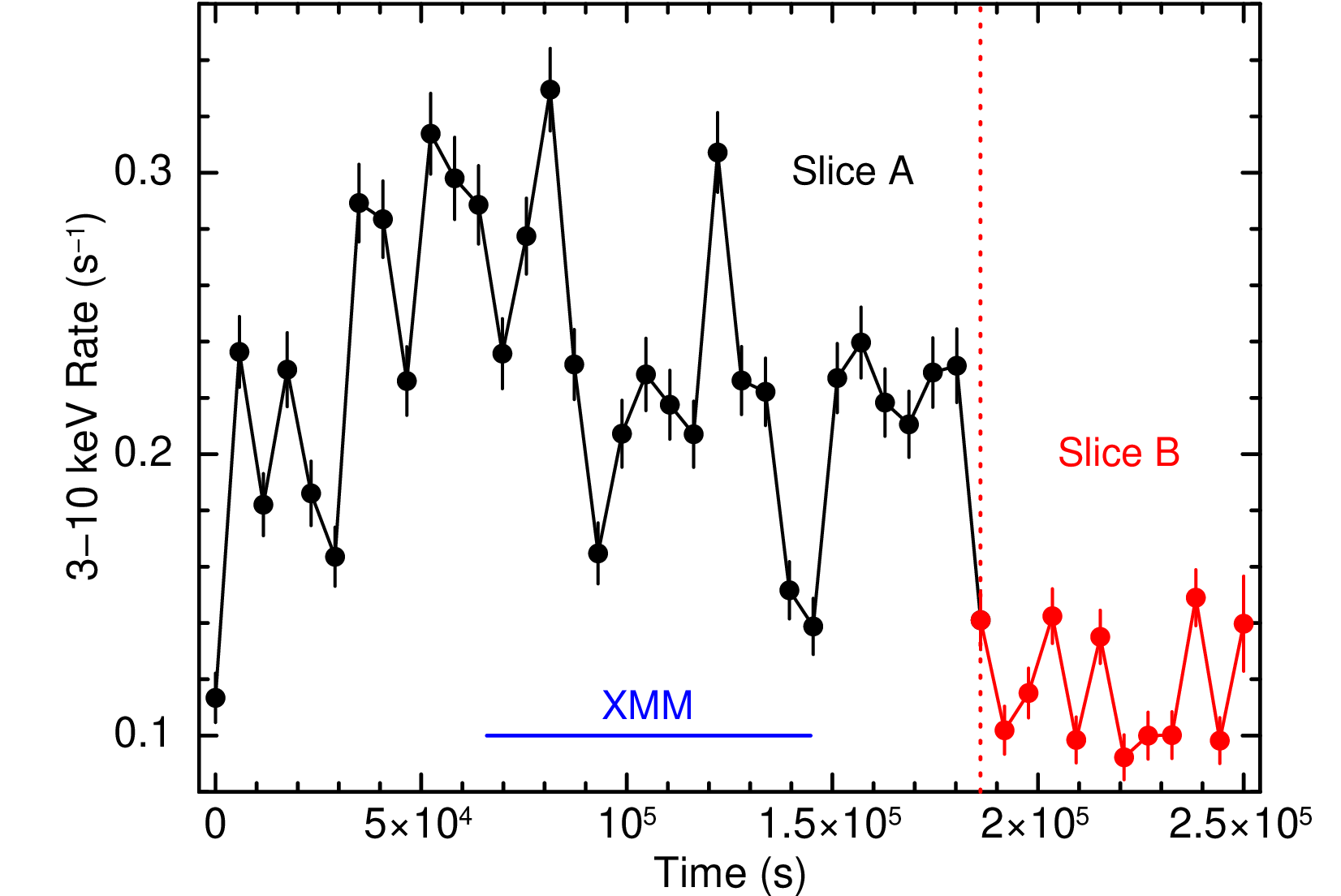}}
\end{center}
\vspace*{-0.5cm}
\caption{Net source lightcurve from the 2022 \pg\ {\it NuSTAR} observation, from 3--10\,keV and over 5814\,s orbital bins, from the FPMA and FPMB detectors combined. The black points correspond to the first 190\,ks when the source was at a higher flux, with the red points occurring during a lower flux period. These intervals (indicated by a red dashed vertical line) were used to define the time intervals for the spectral analysis in Section~4 (slices A and B respectively). The horizontal blue line marks the period when the \xmm\ observation was performed.}
\label{fig:lc} 
\end{figure}

Spectra from the {\it XMM-Newton} Reflection Grating Spectrometer (RGS, \citealt{denHerder01}) for both the 2022 and 2017 epochs were extracted using the \textsc{rgsproc} pipeline. These were combined into a single spectrum for each epoch using using \textsc{rgscombine}, after first checking that the individual 
RGS 1 and RGS 2 spectra were consistent with each other within the errors. The total net count rate obtained over the 6--30\,\AA\ band was $0.336\pm0.002$\,cts\,s$^{-1}$ per detector (75.5\,ks net exposure per RGS) for 2022 and $0.068\pm0.001$\,cts\,s$^{-1}$ per detector (115.8\,ks net exposure per RGS) for 2017. Thus the source count rate is typically a factor of 4--5 lower in the 2017 observation compared to 2022.
The spectra were binned in constant wavelength bins of $\Delta\lambda=0.1$\,\AA, which approximates the spectral resolution of the RGS gratings. Note with this binning, there are typically $\sim50$ and $\sim 200$ counts per 
resolution bin in the 2017 and 2022 RGS spectra respectively.

\section{Variability of \pg}

In this section, the variability properties of \pg\ over the long 250\,ks duration {\it NuSTAR} observation are described.  
A 2022 {\it NuSTAR} lightcurve of \pg\ was initially extracted over the 3--10\,keV band over the full $\sim250$\,ks duration of the observation. The net source lightcurve is shown in the upper panel of Figure\,1, for the FPMA and FPMB modules combined, binned into satellite orbital bins of 5814\,s. 
The source shows prominent variability, by a factor of $\times3-4$ over the minimum to maximum count rate range. 
In particular, a pronounced and rapid decline in the source flux is seen at about 190\,ks into the {\it NuSTAR} observation (Figure~\ref{fig:lc}, red points), with the decrease occurring within two {\it NuSTAR} orbits, or 10\,ks in elapsed time, after which the source flux remains low. The mean 3--10\,keV count rate decreased by a factor of two from $0.213\pm0.003$\,cts\,s$^{-1}$ over the first 190\,ks of the observation to $0.113\pm0.003$\,cts\,s$^{-1}$, 
during the last 190--250\,ks interval. In Section~4 we analyze the spectra taken from the first 0--190\,ks of the observation (hereafter slice A) and the last 60\,ks of the observation (hereafter slice B), where the flux remained low. 
As is seen in Figure~\ref{fig:lc} (blue horizontal line), the \xmm\ observation coincided with the brighter slice A interval of the {\it NuSTAR} observation.

\subsection{Fractional Variability Spectra}

To quantify in greater detail the spectral variability during the {\it NuSTAR} observation, we investigate the fractional rms variability (or $F_{\rm var}$) spectrum \citep{Parker20, Igo20}. 
For the analysis, we extracted {\it NuSTAR} lightcurves over finer energy bins in increments of $\Delta E=0.5$\,keV from 3--10\,keV and from 10--12\,keV, 12--15\,keV and 15--30\,keV over the higher 
energy portion of the 3--30\,keV {\it NuSTAR} bandpass. 
Following the description provided in \citet{Vaughan03}, we then compute the excess variance, $\sigma^{2}_{\rm XS}$, which is defined as: $\sigma^{2}_{\rm XS} = S^{2} - \overline{\sigma^{2}_{\rm err}}$, where $S^{2}$ is the sample variance and $\overline{\sigma^{2}_{\rm err}}$ is the mean square error. We then compute the normalized excess variance by dividing by the squared mean count rate in each band; i.e. $\sigma^{2}_{\rm NXS} = \sigma^{2}_{\rm XS} / \overline{x}^2$. The square root of this value gives the fractional variability, $F_{\rm var}$, which allows us to express the normalized excess variance as a percentage. Errors on $F_{\rm var}$ are given by equation B2 in \citet{Vaughan03}.

\begin{figure}
\begin{center}
\rotatebox{-90}{\includegraphics[height=8cm]{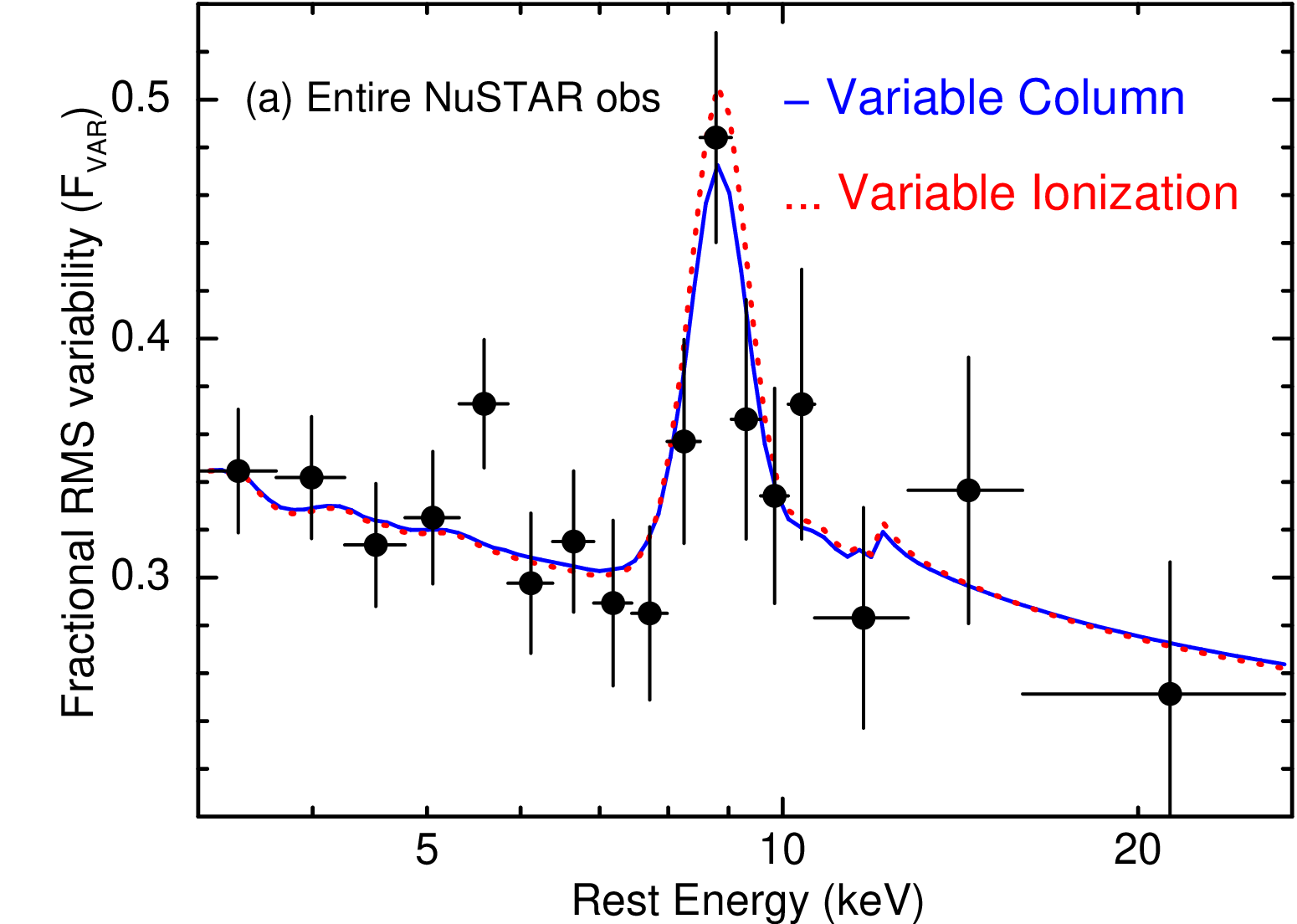}}
\rotatebox{-90}{\includegraphics[height=8cm]{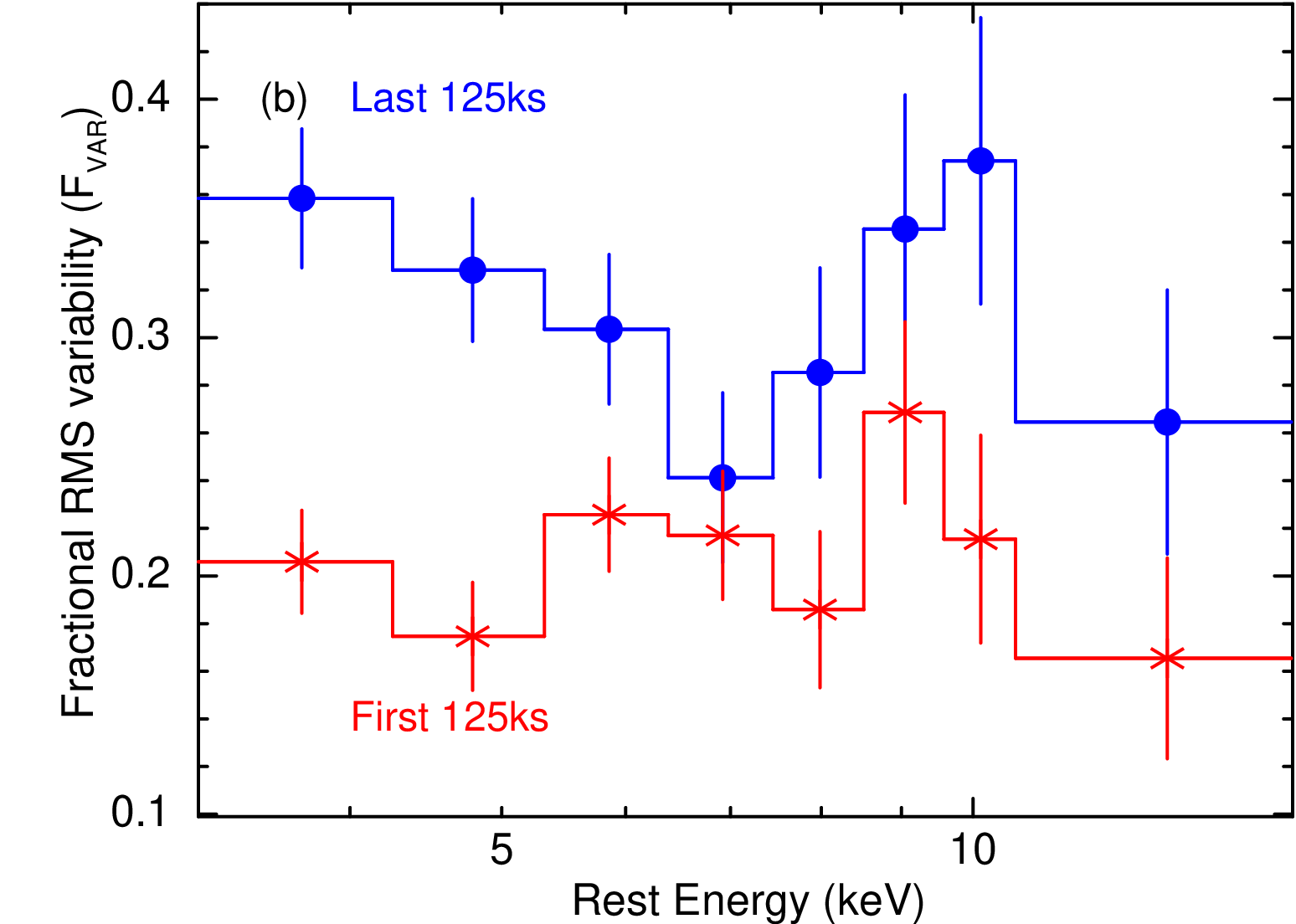}}
\end{center}
\caption{The {\it NuSTAR} fractional variability (or $F_{\rm var}$) spectrum of \pg, calculated from; (a) the entire observation and (b) the first and last 125\,ks respectively. In panel (a), the AGN shows enhanced variability between 8--9\,keV, which may be due to the variability of an Fe K absorption trough from an ultra fast outflow. For comparison, the solid blue line and dotted red lines are the theoretical predictions from an ionized absorber, which is either variable in column or ionization (see Section~4 for details). The lower panel (b) shows the $F_{\rm var}$ spectra from the first and second halves of the observation, split at 125\,ks. The first (bright) part (red curve) shows no energy dependent spectral variability, while the second part of the observation (blue curve, including the flux drop period) displays enhanced variability both in the iron K band between 8--10\,keV and towards the lowest energy bins.}
\label{fig:fvar}
\end{figure}

The resulting $F_{\rm var}$ spectrum of \pg\ over the entire {\it NuSTAR} observation is shown in Figure~\ref{fig:fvar} (upper panel a), transposed into the AGN rest frame at $z=0.0645$. A prominent spike in the $F_{\rm var}$ spectrum is observed between $8-9$\,keV. Using a simple parameterization, a power-law fit gives a poor fit to the $F_{\rm var}$ spectrum ($\chi^2/\nu=24.0/15$), which is improved (to $\chi^2/\nu=8.6/13$) upon the addition of a Gaussian of positive normalization, centered at $8.9\pm0.2$\,keV. Note that the width of the Gaussian line is fixed to $\sigma=0.5$\,keV, corresponding to the width of the $F_{\rm var}$ energy bins. 
As described in \citet{Parker20}, a variable ultra fast outflow can produce spikes or enhancements in variability which show up in $F_{\rm var}$ spectra. Thus when the continuum flux declines and if the absorption line opacity increases -- for instance due to an increase in column, decrease in ionization or an increase in covering fraction -- a positive signal would be revealed in the $F_{\rm var}$ spectrum due to an enhancement of variability in the energy band where the absorption line is observed. This hypothesis is investigated further in the spectral analysis in Section~4.1.  

To assess any change in the spectral variability over the course of the observation, the $F_{\rm var}$ spectra were re-extracted covering the first half (0--125\,ks) and last half (125--250\,ks) of the observation. A coarser energy binning is adopted (e.g. $\Delta E=1$\,keV between 3--10\,keV) due to the shorter time intervals covered, which limits the signal to noise in the spectra. These two $F_{\rm var}$ spectra are plotted in panel (b) of Figure~\ref{fig:fvar}. It is clear that, in the second half of the observation which includes the low flux period, the overall 
AGN variability is enhanced; e.g. assuming a constant amplitude of variability between the two observation halves results in a very poor description of the spectra $\chi^2/\nu=59.6/16$. 
 Furthermore, in the second half, an excess is observed between 8-10\,keV, while the variability is also enhanced in the softer 3--5\,keV band. In contrast, there appears to be no spectral variability during the first half of the observation, which can be simply fitted with a constant $F_{\rm var}$ versus energy ($\chi^2/\nu=7.5/8$). Thus the spectral variability, possibly originating from a variable absorber or outflow, is restricted to the last 
part of the {\it NuSTAR} observation and is likely to be coincident with the slice B interval. This is now directly tested in the next Section.

\section{Epoch Dependent Spectral Analysis}

\subsection{The 2022 {\it NuSTAR} Observation.}

\begin{figure}
\begin{center}
\rotatebox{-90}{\includegraphics[height=8cm]{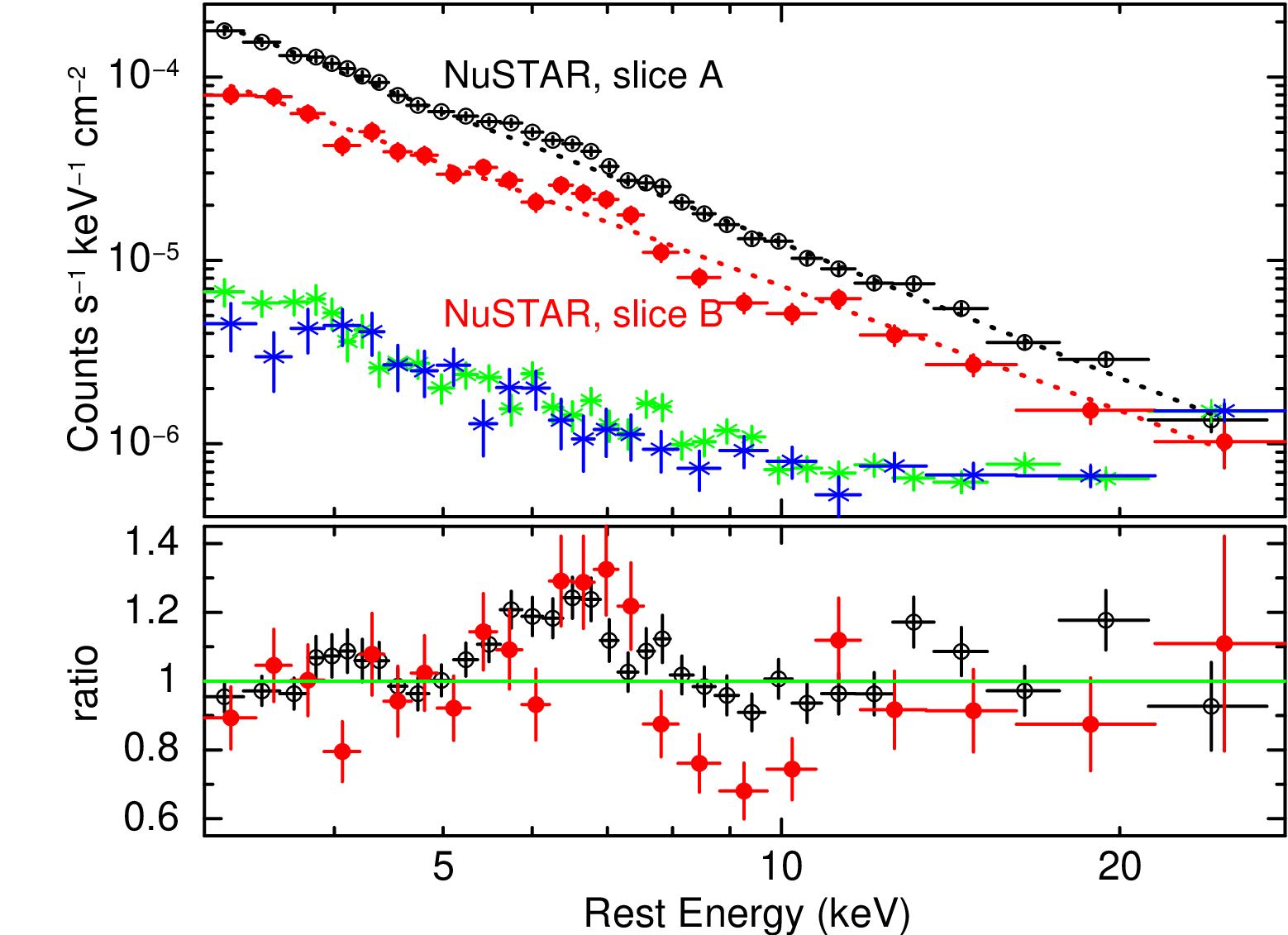}}
\end{center}
\caption{2022 {\it NuSTAR} spectra of \pg, showing the slice A spectrum (black, open circles) versus the slice B spectrum (red, filled circles). 
The upper panel shows the net count rate spectra for both observations, while the background spectra for the slice A and B segments are shown in green and blue respectively 
and the dotted lines show a simple power-law continuum. 
The lower panel shows the ratio of the above power-law to both spectra. This reveals an excess of emission near 6 keV in both spectra and a strong deficit of counts between 8--10\,keV, 
which is apparent only in the slice B low flux spectrum. This may be due to the emergence of an ultra fast outflow in the last 60\,ks during the low flux period.}
\label{fig:nustar}
\end{figure}

Next, the aim is to perform a simple quantitative comparison of the iron K-shell region during the {\it NuSTAR} slice A and B intervals. In particular this allow us to assess the presence of a variable absorption trough, as indicated by the $F_{\rm var}$ analysis above. 
The full useable 3--30\,keV {\it NuSTAR} band was used for this purpose. 
Spectra were thus extracted from slices A (0--190\,ks) and B (190--250\,ks) intervals, as indicated in Figure\,1, where slice B is obtained purely from the flux drop interval. The net source rates (exposure times) or the two intervals are:- $0.1341\pm0.0012$\,cts\,s$^{-1}$ (97.3\,ks)  and $0.0682\pm0.0016$\,cts\,s$^{-1}$ (33.1\,ks), for 
slice A and B respectively. The 3--30\,keV band {\it NuSTAR} count rate spectra are shown in Figure~\ref{fig:nustar} (left panel) and the spectra for both intervals lie well above the background level below 30\,keV. The 3--30\,keV band source flux drops by a factor of two from $F_{3-30}=5.0\times10^{-12}$\,ergs\,cm$^{-2}$\,s$^{-1}$ (slice A) to 
$F_{3-30}=2.6\times10^{-12}$\,ergs\,cm$^{-2}$\,s$^{-1}$ (slice B). 

\begin{deluxetable}{lll}
\tablecaption{{\it NuSTAR} Iron K profile parameters for PG\,1448+273.}
\tablewidth{250pt}
\tablehead{
& \colhead{SLICE\,A} & \colhead{SLICE\,B}}
\startdata
Gaussian emission:-\\
$E_{\rm rest}$\,(keV) & $6.22\pm0.20$ & $6.83\pm0.25$ \\
$\sigma$\,(keV) & $0.63^{+0.33}_{-0.19}$ & $0.37^{+0.30}_{-0.17}$\\
Line Flux\,($\times10^{-5}$\,photons\,cm$^{-2}$\,s$^{-1}$) & $1.6\pm0.5$ &  $0.9\pm0.4$ \\
EW\,(eV)& $320\pm120$ &  $440\pm190$ \\
\hline
Gaussian absorption:-\\
$E_{\rm rest}$\,(keV) & $9.2\pm0.3^t$ & $9.2\pm0.3$ \\
$\sigma$\,(keV) & $0.78^{+0.32t}_{-0.26}$ & $0.78^{+0.32}_{-0.26}$\\
Line Flux\,($\times10^{-5}$\,photons\,cm$^{-2}$\,s$^{-1}$) & $|<0.63|$ &  $-0.80\pm0.27$ \\
EW\,(eV) & $|<210|$ &  $-750\pm250$ \\
\hline
Continuum:-\\
$\Gamma$ & $2.44\pm0.04$ & $2.32\pm0.09$\\ 
$F_{\rm 3-30\,keV}$\,($\times10^{-12}$\,ergs\,cm$^{-2}$\,s$^{-1}$) & 5.0 & 2.6\\
\hline
Statistics:-\\
$\chi_{\nu}^{2}$ (PL only) & $165.1/103$ & $58.9/22$\\
$\chi_{\nu}^{2}$ (with lines) & $102.7/100$ & $16.0/16$
\enddata
\tablenotetext{t}{Denotes parameter is tied in fit.}
\label{tab:profile}
\end{deluxetable}

Figure~\ref{fig:nustar} shows that significant residuals are apparent in the iron K band versus a simple steep ($\Gamma\approx2.4$) power-law model, which results in a poor fit to both spectra 
($\chi^2/\nu=165.1/103$, slice A; $\chi^2/\nu=58.9/22$, slice B).  Note that a neutral Galactic column of $N_{\rm H}=3.0\times10^{20}$\,cm$^{-2}$ \citep{Kalberla05} was also included. 
An excess of emission is observed in the data/model residuals to both spectra between 5--7\,keV, however only slice B shows a strong absorption trough which is present in the residuals between 8--10\,keV. 

To provide a simple quantification of the profile, the emission and absorption were parameterized with simple Gaussians with variable energy, width and normalization (photon flux) and where the normalization of the absorption line is set to a negative value. 
The results are summarized in Table~\ref{tab:profile}. In the slice A spectrum, the addition of a broad Gaussian emission line significantly improved the fit to an acceptable value of $\chi^2/\nu=102.7/100$, with no requirement for any additional absorption. In contrast, the slice B spectrum requires both excess absorption and emission; e.g. adding an absorption line to the power-law continuum improved the fit to $\chi^2/\nu=34.9/19$, which decreased further to $\chi^2/\nu=16.0/16$ upon the addition of an iron K emission line. Here, the absorption line has a rest frame energy of $9.3\pm0.3$\,keV, a width of $\sigma=0.78^{+0.32}_{-0.26}$\,keV and an equivalent width of $-750\pm250$\,eV. The line energy is significantly blue-shifted compared to the expected lab-frame energy of either Fe\,\textsc{xxv} He$\alpha$ at 6.70\,keV or Fe\,\textsc{xxvi} Ly$\alpha$ at 6.97\,keV, with corresponding relativistically corrected blueshifts of $v/c=-0.31\pm0.03$ or  $v/c=-0.27\pm0.03$ respectively. The 9\,keV absorption trough found in the slice B interval also occurs at the same energy as the spike in the above $F_{\rm var}$ spectrum. In contrast, only an upper-limit is found on the equivalent width of the absorption line in slice A, of ${\rm EW}<210$\,eV for the same line energy and width as above. 

\subsubsection{Photoionization Modelling} \label{subsec:photoionization_modelling}

In order to obtain a simple parameterization of the ionized gas and its variability with respect to the continuum, we then modeled the {\it NuSTAR} spectra with a self consistent photoionization model,
using the \textsc{xstar} code (\citealt{Kallman04}). 
The absorption was accounted for by a multiplicative grid, while the emission from the wind was modeled by a Gaussian emission line. 
The absorber was initially assumed to fully cover the power-law and iron K emission component. 
A velocity broadening of $b=25\,000$\,km\,s$^{-1}$ was used in the models, accounted for by the turbulence velocity parameter and is consistent with the line widths inferred from the earlier Gaussian analysis.
Solar abundances of \citet{GrevesseSauval98} were used throughout. 
Here we use a $\Gamma=2.4$ power-law continuum between 1--1000 Rydberg, as is observed in the 2022 {\it NuSTAR} spectra out to 30\,keV.


The slice A and B spectra were fitted simultaneously considering two possible scenarios; (i) where the column density is allowed to vary between the datasets (with constant ionization) and (ii) 
where the ionization varies and the column is constant. The results are shown in Table~\ref{tab:xstar} and both scenarios give an equally good fit to the {\it NuSTAR} spectra. The increase in opacity at iron K 
can either be accounted for by an increase in column from $N_{\rm H}<1.9\times10^{23}$\,cm$^{-2}$ to $N_{\rm H}=7.2^{+3.6}_{-3.2}\times10^{23}$\,cm$^{-2}$ or a decrease in ionization from $\log\xi=5.6^{+0.5}_{-0.2}$ to 
$\log\xi=4.9\pm0.3$, from slice A to slice B respectively. 

Alternatively, instead of a change in column density, the variable depth of the 9\,keV absorption trough might be accounted for by a change in the line of sight covering fraction of the absorber. 
To test this, a constant column density was assumed (adopting the above best fit value of $N_{\rm H}=7.2\times10^{23}$\,cm$^{-2}$) along with a constant ionization parameter of $\log \xi=4.9$. 
Slice B containing the deep absorption trough was assumed to have a 100\% covering fraction, while this constraint was relaxed for slice A containing the shallow absorption. 
In this case, a lower covering fraction of $30^{+19}_{-17}$\% can account for the weaker absorption in the slice A spectrum. Thus within the limits of these data, a variable covering fraction cannot easily be distinguished 
from an intrinsically variable column density and both can replicate changes in the profile depth. 

\begin{deluxetable*}{lcccc}
\tablecaption{Photoionization Modeling of the Fe K Wind.}
\tablewidth{500pt}
\tablehead{\colhead{Parameter} & \colhead{{\it NuSTAR}/SLICE\,A} & \colhead{{\it NuSTAR}/SLICE\,B} & \colhead{XMM/2017} & \colhead{XMM/2022}}
\startdata
Energy range (keV) & 3--30\,keV & 3--30\,keV & 3--10\,keV & 3--10\,keV \\
\hline
Variable $N_{\rm H}$:-\\
$N_{\rm H}$ ($\times10^{23}$\,cm$^{-2}$) & $<1.9$ & $7.2^{+3.6}_{-3.2}$ & $6.1\pm1.2$ & $<0.9$\\
$\log(\xi/{\rm ergs\,cm\,s}^{-1})$ & $4.9\pm0.3$ & $4.9\pm0.3^t$ & $4.9\pm0.2$ & $4.9\pm0.2^t$\\
$v/c$  & $-0.30\pm0.03^t$ & $0.30\pm0.03$ & $-0.086\pm0.005$  & $-0.086\pm0.005^t$\\
$\chi^{2}/\nu$ & 103.9/99 & 18.7/17 & 213.1/210 & 150.5/161\\
\hline
Variable $\log\xi$:-\\
$N_{\rm H}$ ($\times10^{23}$\,cm$^{-2}$) & $7.2^f$ & $7.2^f$ & $6.1\pm1.2$ & $6.1\pm1.2$\\
$\log(\xi/{\rm ergs\,cm\,s}^{-1})$ & $5.6^{+0.5}_{-0.2}$ & $4.9\pm0.3$ & $4.9\pm0.2$ & $>5.8$\\
$v/c$  & $-0.29\pm0.03^t$ & $0.29\pm0.03$ & $-0.087\pm0.005$ & $-0.087\pm0.005^t$\\
$\chi^{2}/\nu$ & 103.9/99 & 18.4/18 & 213.7/210 & 150.2/161\\
\hline
Fe K Emission:-\\
$E_{\rm rest}$\, (keV) & $6.21\pm0.20$ & $6.81\pm0.21$ & $6.48^{+0.34t}_{-0.24}$ & $6.48^{+0.34}_{-0.24}$\\
$\sigma$\, (keV) & $0.60^{+0.30}_{-0.19}$ & $0.35^{+0.22}_{-0.17}$ & $0.65^{+0.35t}_{-0.25}$ & $0.65^{+0.35}_{-0.25}$\\
Line Flux ($\times 10^{-5}$\,photons\,cm$^{-2}$\,s$^{-1}$) & $1.5\pm0.4$ &  $0.78\pm0.32$ & $<0.46$ & $0.85\pm0.36$\\
EW\,(eV) & $300\pm80$ &  $370\pm150$ & $<230$ & $215\pm90$\\
\hline
Continuum:-\\
$\Gamma$ & $2.45\pm0.04$ & $2.32\pm0.09$ & $1.85\pm0.06$ & $2.39\pm0.05$ \\
$F_{\rm 3-10\,keV}$\, ($\times10^{-12}$\,erg\,cm$^{-2}$\,s$^{-1}$) & 3.23 & 1.60 & 1.12 & 2.84
\enddata
\tablenotetext{f}{Denotes parameter is fixed.}
\tablenotetext{t}{Denotes parameter is tied between observations.}
\label{tab:xstar}
\end{deluxetable*}

Following the above results, the $F_{\rm var}$ spectrum of \pg\ in Section~3.1 was also revisited. To calculate the contribution of the absorber variability to the {\it NuSTAR} $F_{\rm var}$ spectrum, a variable multiplicative 
component was computed and applied to the $F_{\rm var}$ spectrum. This was calculated assuming either case (i) above, where the column density varies by $\Delta N_{\rm H}=7\times10^{23}$\,cm$^{-2}$ (for a constant ionization parameter of 
$\log\xi=4.9$), or (ii) where the ionization parameter varies by $\Delta\log\xi=0.7$ (for a constant column of $N_{\rm H}=7\times10^{23}$\,cm$^{-2}$). In each case, an absorber multiplicative factor is calculated as a function of energy, which can then be subsequently multiplied by the underlying continuum variability, which is modeled by a power-law in the $F_{\rm var}$ spectrum. The result of applying this variability model to the $F_{\rm var}$ spectrum is shown in the upper panel of Figure~\ref{fig:fvar}, whereby the solid blue line presents the variable $N_{\rm H}$ case and the dotted red line the variable $\log\xi$ case. Both cases can reproduce the observed $F_{\rm var}$ spectrum, with an identical fit statistic of $\chi^2/\nu=8.1/13$, 
where the spike at 9\,keV results from an increase in absorber opacity in the low flux spectrum. 
The same is also true for the variable covering fraction, which yields an identical result to the variable column case.  
Thus the change in opacity from slice A to B can originate from either a change in column, ionization or covering of a fast wind, which accounts for both the spectral changes and the $F_{\rm var}$ behavior. 

\subsection{Comparison between the 2017 and 2022 epochs}

\begin{figure}
\begin{center}
\rotatebox{-90}{\includegraphics[height=8cm]{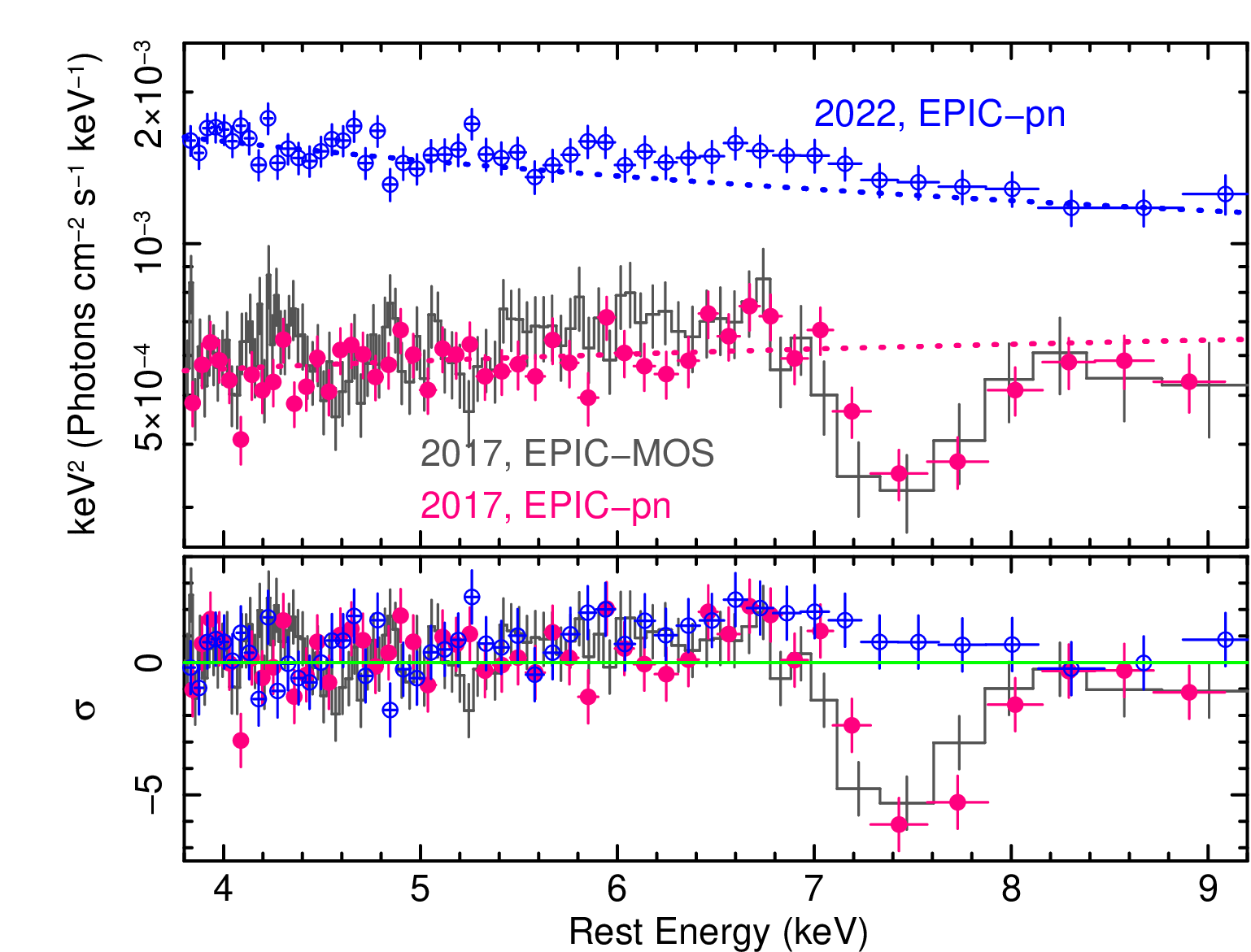}}
\end{center}
\caption{The 2017 \xmm\ EPIC-pn (red) and MOS (grey) spectra of \pg, compared to the bright state 2022 EPIC-pn spectrum (blue). The 2017 spectra are a factor of three fainter in the 2--10\,keV band and have a flatter photon index of $\Gamma=1.9$ versus the steeper $\Gamma=2.4$ continuum observed in 2022, where the respective power-laws are shown by dotted lines. The lower panel shows the residuals in units of $\sigma$ compared to a simple power-law. A significant broad absorption line, centered near 7.5\,keV, is present in both the 2017 pn and MOS spectra, but is not present in the higher flux 2022 spectrum. The derived outflow velocity in 2017 is three times lower than the fast ($0.3c$) wind observed in the 2022 {\it NuSTAR} slice B interval and where the absorption trough is instead centered near 9\,keV.} 
\label{fig:xmm}
\end{figure}

The 2017 and 2022 \xmm\ spectra of \pg\ were also compared to assess the long term variability of the wind, where the spectra are plotted in Figure~\ref{fig:xmm}. The \xmm\ data were initially limited to the 3--10\,keV band to parameterize 
the iron K emission and absorption and to provide a direct comparison to the hard X-ray {\it NuSTAR} spectra. The broad-band multi-epoch spectra, which contain a soft excess and warm absorber below 3\,keV, will be described in detail in Sections 5 and 6.

As expected given its simultaneity, the 2022 \xmm\ spectrum is very similar in properties to the slice A {\it NuSTAR} spectrum, showing a steep power-law spectrum ($\Gamma=2.39\pm0.05$), a weak broad iron K emission line and no evidence of any iron K absorption trough in the 7--10\,keV band. The 3--10\,keV flux of the 2017 spectrum is lower by a factor of three compared to 2022 (see Table~\ref{tab:xstar}). As reported by \citet{Kosec20} and \citet{Laurenti21}, a strong broad absorption trough is apparent with a centroid energy of between 7.4--7.5\,keV. From fitting a simple Gaussian profile, the rest frame energy centroid is 
$E=7.47\pm0.07$\,keV, with a width of $\sigma=0.30^{+0.09}_{-0.07}$\,keV and an equivalent width of ${\rm EW}=-390\pm75$\,eV. The addition of the Gaussian absorption profile in 2017 is highly significant, with 
the fit statistic decreasing from $\chi^2/\nu=351.7/215$, for a power-law fit, to $\chi^2/\nu=222.9/212$, upon the inclusion of the absorption line in the model. 
In contrast, there is no absorption present in the brighter 2022 \xmm\ spectrum, with an upper limit of ${\rm EW}<125$\,eV for the same line energy and width as above. 

There are two notable differences between the Fe K absorption profile observed in the 2017 \xmm\ low flux spectrum compared to what is seen in the {\it NuSTAR} slice B low flux spectrum. Firstly, the line energy is substantially lower in the 2017 spectrum (2017, $E=7.47\pm0.07$\,keV versus 2022 slice B, $E=9.2\pm0.3$\,keV). 
Secondly, the line width is at least a factor of two lower in 2017 compared to the 2022 slice B spectrum ($\sigma=0.30^{+0.09}_{-0.07}$\,keV vs $0.78^{+0.32}_{-0.26}$\,keV). This points to a lower velocity wind being present in the 2017 epoch.

\begin{figure*}
\begin{center}
\rotatebox{-90}{\includegraphics[height=8cm]{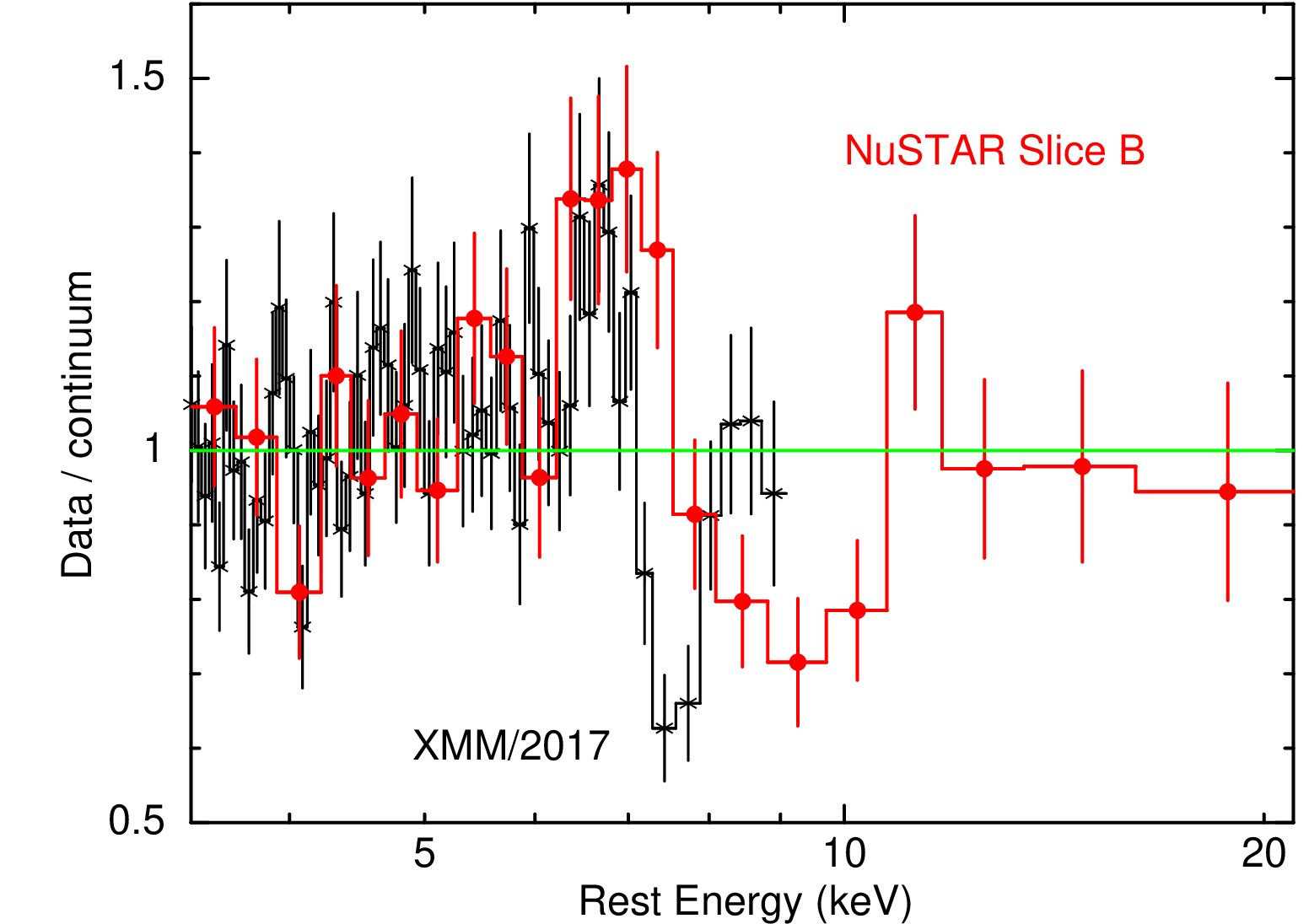}}
\rotatebox{-90}{\includegraphics[height=8cm]{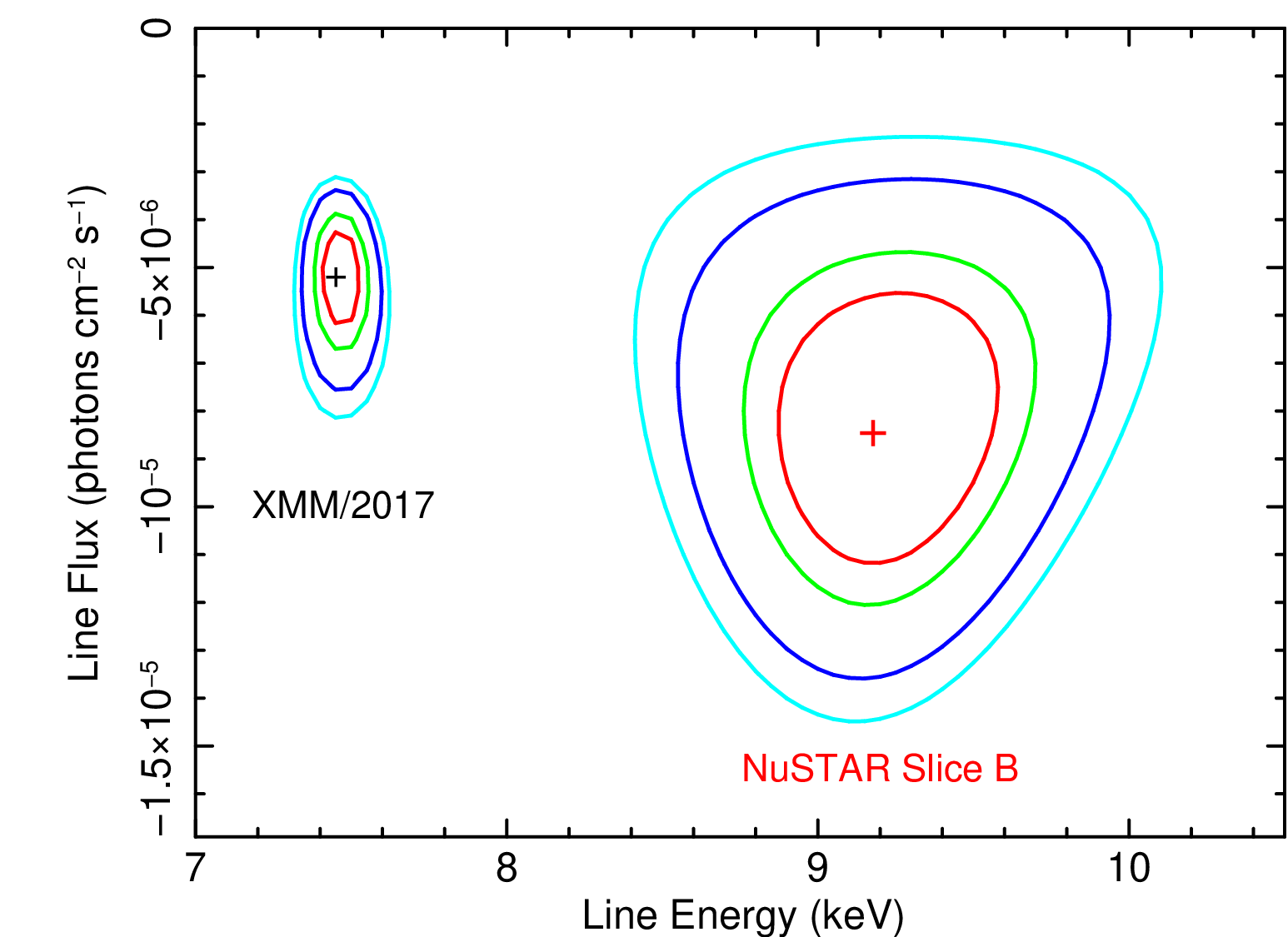}}
\end{center}
\caption{A detailed comparison between the Fe K line profiles between the 2017 \xmm\ (EPIC-pn, black points) and 2022 slice B ({\it NuSTAR}, red points) spectra. The left panel shows the overlay of the two profiles, plotted as a ratio to a power-law continuum. The low energy trough in the 2017 \xmm\ spectrum occurs well before the onset of the high energy absorption trough in 2022 slice B, which is located between 8--10\,keV. The right hand panel shows the 68\%, 90\%, 99\% and 99.9\% confidence contours between the absorption line centroid energy and flux for each profile. The slice B profile shows a significantly higher blue-shift compared to 2017, at $>99.9$\% confidence.} 
\label{fig:compare}
\end{figure*}

These differences can be seen in Figure~\ref{fig:compare}. The left panel shows an overlay of the data/model ratio to a power-law continuum to both the 2017 \xmm\ and the 2022 slice B {\it NuSTAR} spectra, while the right hand panel shows the confidence contours on the line energy versus flux for each dataset with a clear separation in line energy, at $>99.9$\% confidence. 
From comparing the two profiles, the 7.5\,keV absorption line in 2017 occurs before the onset of the 2022 slice B broad absorption trough, which is 
confined to between 8--10\,keV. The limit on a Gaussian absorption line in the slice B spectrum, at the same energy and width as the 2017 one, is ${\rm EW} < 80$\,eV. Thus we can rule out the detection of the slower 
wind profile in the 2022 data and require that its equivalent width decreases by at least a factor of 4, compared to the 2017 profile.

To model the absorber properties, the \textsc{xstar} grid of models was applied to the 2017 and 2022 \xmm\ spectra and the results are listed next to the 2022 {\it NuSTAR} slice A and B epochs in Table~\ref{tab:xstar} for 
direct comparison. One difference for the 2017 epoch is that a lower turbulence velocity of 10000\,km\,s$^{-1}$ was adopted to account for the lower velocity width during this observation. The 2017 and 2022 \xmm\ spectra were then directly compared, allowing either the absorber column or ionization to vary between them to account for the opacity change at 7.5 keV, as per Section~4.1.2. Similar to the previous \textsc{xstar} analysis, either a change in column ($\Delta N_{\rm H}=6\times10^{23}$\,cm$^{-2}$) or ionization (by $\Delta\log\xi=0.9$) or covering fraction can account for the difference between the spectra, in addition to the variable power-law continuum. The low column (or high ionization) derived from the 2022 \xmm\ spectrum results from the lack of an iron K absorption trough at 7.5\,keV. This is also similar to what was found by \citet{Kosec20} and \citet{Laurenti21} in a short 20\,ks \xmm\ snapshot of \pg\ in 2003 and which was also at a higher flux than in 2017, where no fast wind component was required in that epoch. 

Furthermore, while the wind column density in 2017 ($N_{\rm H}=6.1\pm1.2\times10^{23}$\,cm$^{-2}$, Table~\ref{tab:xstar}) is similar to what is observed in the 2022 slice B {\it NuSTAR} spectrum, it is clear that the outflow velocity, of $v/c=-0.087\pm0.005$, is lower by a factor of three compared to slice B, where $v/c=-0.29\pm0.03$. Thus the wind appears to have either significantly changed in its velocity between the 2017 and 2022 epochs, or that we are viewing different components of the wind at different times. 
With respect to the latter, as there was no joint {\it NuSTAR} observation of \pg\ in 2017, it is not possible to probe whether the higher velocity ($v/c\sim0.3$) wind component is present in addition to the slower zone. Nonetheless, the slower 2017 wind component is not required in the higher flux 2022 \xmm\ observation, while the low flux 2022 slice B interval is dominated by a much faster wind component with $v/c=0.3$ with no requirement for a slower zone.


\section{Disk Wind Modeling}

\subsection{The Disk Wind Parameters}

In order to model the wind signatures in the \pg\ spectra in a more physical context and to provide estimates of the wind parameters, 
we utilized the radiative transfer disk wind code developed by \citet{Sim08,Sim10a}. The analysis was performed over the broad 0.3--30\,keV band. 
The model has been previously employed to fit the Fe K wind absorption profiles in several AGN, e.g. Mrk 766 (\citealt{Sim08}), 
PG\,1211+143 (\citealt{Sim10a}), PDS\,456 (\citealt{Reeves14}), \zw\ \citep{ReevesBraito19}, MCG\,$-$03$-$58$-$007 \citep{Braito22}. 
Recently, \citet{Matzeu22} expanded the parameter ranges covered by this wind model and tested the resulting grids on the 
prototype example of a fast disk wind in PDS\,456. Here, their \textsc{fast32} grid of disk wind models is employed, where the parameters are 
described in detail by \citet{Matzeu22}\footnote{The model is also available at https://gabrielematzeu.com/disk-wind/}. The key parameters of this model are summarized below.

\begin{itemize}
\item Launch radius. The launch radii adopted in this grid are $R_{\rm min}=32R_{\rm g}$ and $R_{\rm max}=48R_{\rm g}$ (where $R_{\rm g}$ is the gravitational radius). 
\item Opening angle. This is set to $\theta=\pm45\degg$ with respect 
to the polar ($z$) axis.
\item Terminal velocity. 
The terminal velocities ($v_{\infty}$) realized in the wind models are determined via
\begin{equation}
v_{\infty} = f_{\rm v} \sqrt{2GM_{\rm BH}/R_{\rm min}}.
\end{equation}
The terminal velocity is adjusted by varying the $f_{\rm v}$ parameter, for a given launch radius (here $R_{\rm min}=32R_{\rm g}$). 
Thus for $f_{\rm v}=1$, the terminal velocity is $0.25c$. 
The \textsc{fast32} grid, covers 8 velocity values, ranging from $f_{\rm v}=0.25-2.0$ (or $v_{\infty}=0.0625-0.50c$).

\item Input Continuum. This was set to be a power-law, covering the range from $\Gamma=1.6-2.4$, in $\Delta\Gamma=0.2$ increments.

\item Inclination angle. The observer's inclination is defined as $\mu = \cos\theta$, where $0.025<\mu<0.975$ over 20 incremental values (with $\Delta\mu=0.05$). Here, $\theta$ is the angle between the observer's line-of-sight and the polar $z$ axis of the wind, with the disk lying in the $xy$ plane (see Figure~1, \citealt{Matzeu22}). 

\item Mass outflow rate. This is defined by the ratio $\dot{M}=\dot{M}_{\rm out}/\dot{M}_{\rm Edd}$, where the mass outflow rate is normalized to 
the Eddington value. 
The grid of models was generated covering the range $\dot{M}=0.02-0.68$, in $n=12$ increments, with equal logarithmic spacing. 

\item Ionizing X-ray luminosity. The X-ray luminosity is parameterized in the 2-10\,keV band as a percentage of the Eddington luminosity, 
where $L_{\rm X} = L_{\rm 2-10 keV}/L_{\rm Edd}$. 
The \textsc{fast32} grid covers the range of $L_{\rm X}$ of 0.025\% to 2.5\% of the Eddington luminosity, over $n=9$ increments in equal logarithmic spacing. 
\end{itemize} 

In order to calculate the total mass outflow rate, the model assumes that the wind is axisymmetric about the azimuthal direction. 
The absorption against the direct continuum arises purely along the line of sight, while the emission from the wind is integrated from photons scattered over the full range of angles above the plane of the disk. 
Therefore, the combination of both the wind emission, in the form of the broad iron K$\alpha$ line and the absorption along the line of sight, are important 
in determining the total mass outflow rate. Indeed, this allowed a realistic estimate of the mass outflow rate for the prototype disk wind case of PDS 456, via both its wind emission and absorption \citep{Nardini15}.


\subsection{Application to the Multi Epoch Spectra}

\begin{figure}
\begin{center}
\rotatebox{-90}{\includegraphics[height=8.5cm]{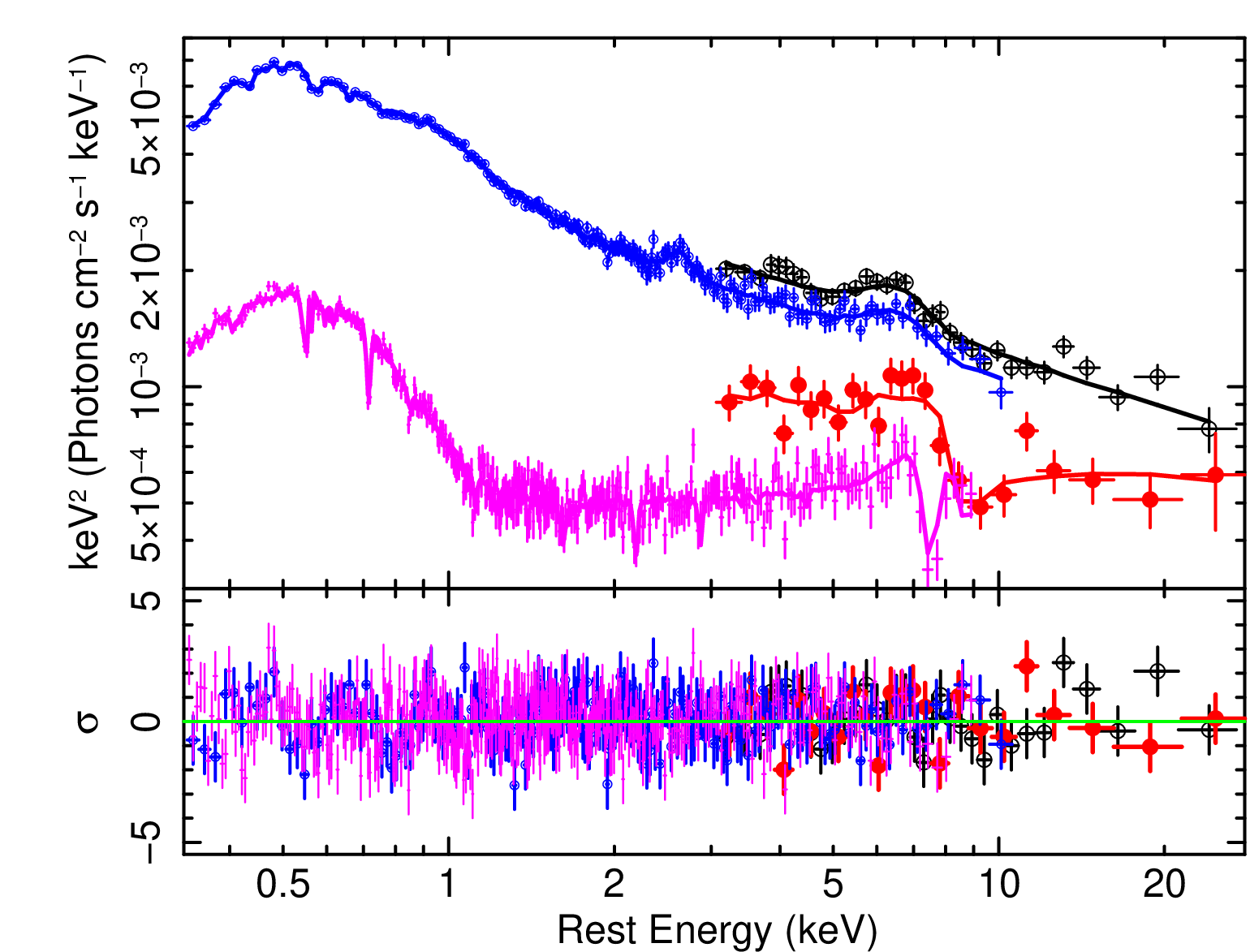}}
\end{center}
\caption{Broad-band X-ray spectra of \pg, fitted with the disk wind model, as listed in Table~4. The {\it NuSTAR} slice A and 2022 EPIC-pn spectra are shown as open circles (black and blue points respectively). 
The slice B spectrum is shown as filled red circles and the 2017 EPIC-pn spectrum is shown as magenta crosses. The EPIC-MOS spectra from 2017 and 2022 are not included in the plot, but are included in the spectral fit. The solid lines shows the best fit model in each case and the lower panel the residuals against this. The disk wind model is able to reproduce the strong iron K absorption profiles present in the 
slice B and 2017 spectra, leaving no significant residuals. Note the soft X-ray \xmm\ spectra below 1\,keV are dominated by a strong soft X-ray excess, especially for the 2017 dataset.}
\label{fig:wind}
\end{figure}

The above disk wind model was then applied to both the slice A and B {\it NuSTAR} spectra, as well as the 2017 and 2022 \xmm\ spectra. 
As the slice A {\it NuSTAR} spectrum overlaps the 2022 \xmm\ spectrum, these were treated as a single epoch (slice A) with identical parameters, 
just allowing for a constant multiplicative factor between the spectra to account for any difference in normalization. 
The slice B {\it NuSTAR} spectrum and the 2017 \xmm\ spectrum were treated as separate epochs, allowing the wind parameters to vary, due to their very different properties. 
Thus in total there are three independent epochs (slice A, slice B and 2017) covering three distinct spectral states of \pg. 
The \xmm\ spectra were modeled over the full 0.3--10\,keV band. Both the 2017 and 2022 \xmm\ spectra require a soft X-ray excess below 2\,keV, 
which was modeled with the thermal Comptonization model \textsc{comptt} \citep{Titarchuk94}. In addition to the Galactic absorption column (of $3\times10^{20}$\,cm$^{-2}$), 
modeled by the \textsc{tbabs} model \citep{Wilms00},  
a soft X-ray warm absorption component (\textsc{wa}) was also included to model the \xmm\ data below 2\,keV, whereby the parameters are fixed to the values determined in the high resolution RGS spectra. 
This will be described in greater detail in Section~6. 
A broad Gaussian emission line (\textsc{gau}) was retained to model any additional iron K emission that is not already accounted for by the disk wind model. 
Thus the overall phenomenological form of the model is:- 
\begin{equation}
\textsc{tbabs}\times\textsc{wa}\times\textsc{wind}\times(\textsc{powerlaw} + \textsc{gau} + \textsc{comptt}).
\end{equation}

\begin{figure}
\begin{center}
\rotatebox{-90}{\includegraphics[height=8cm]{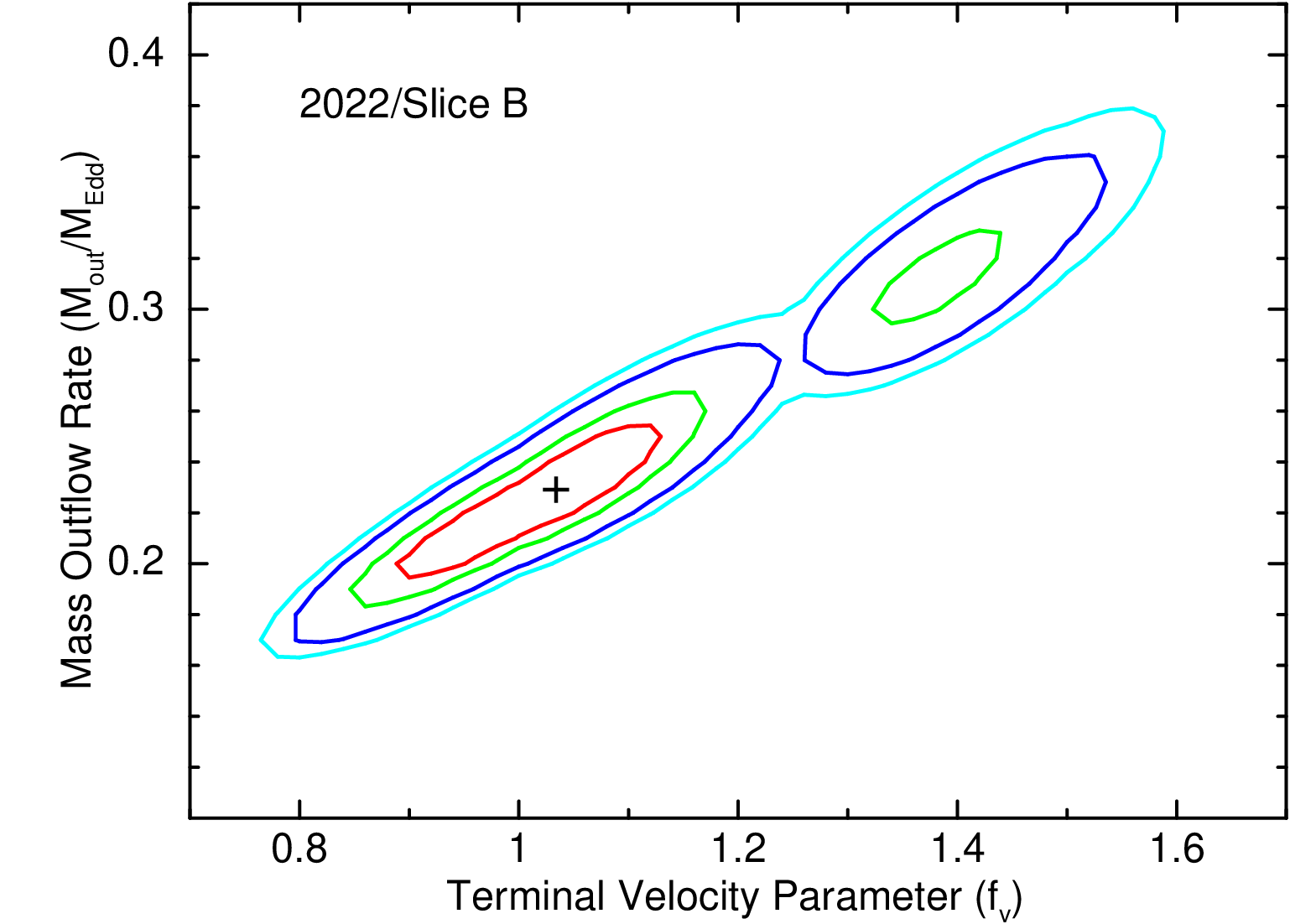}}
\rotatebox{-90}{\includegraphics[height=8cm]{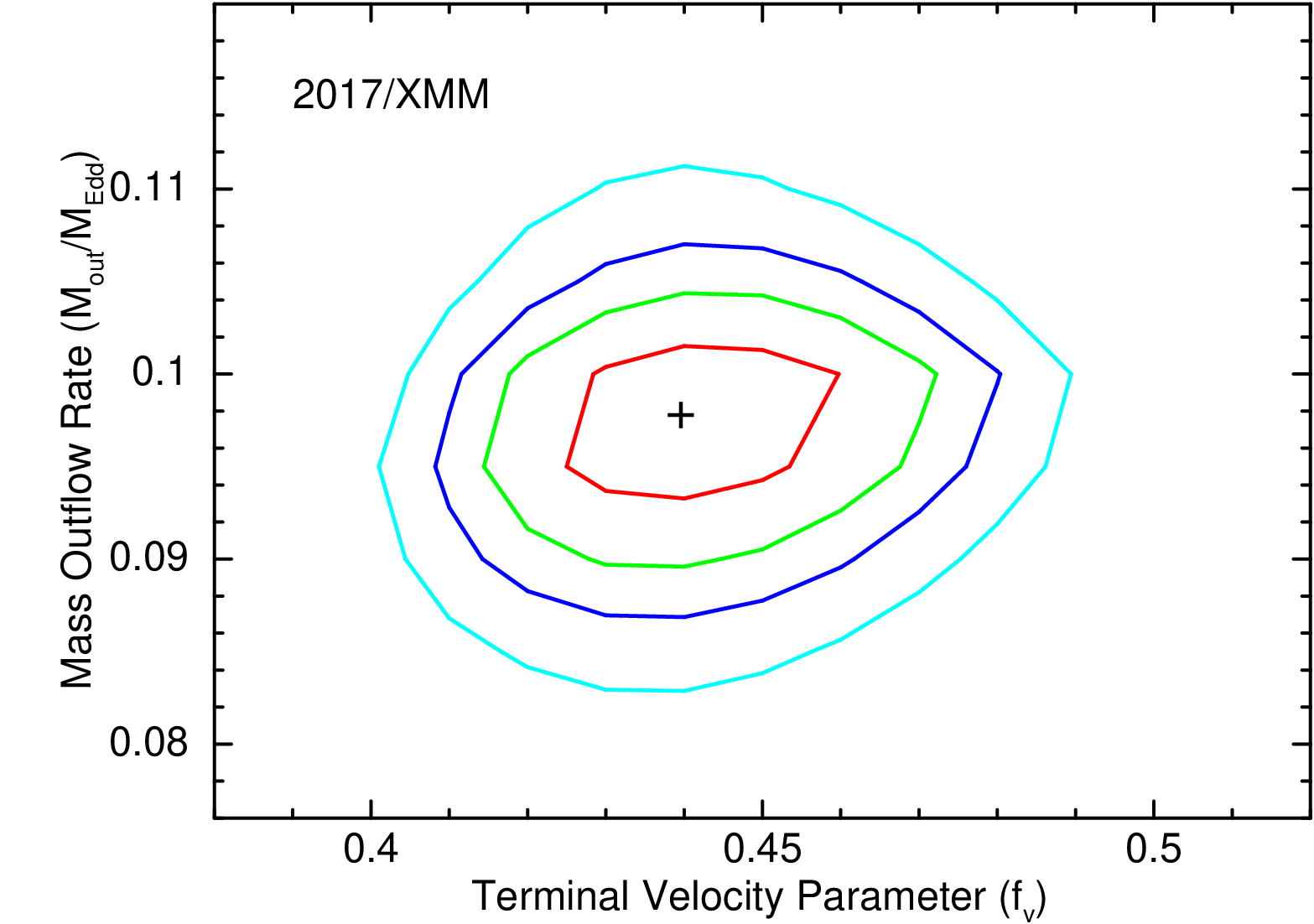}}
\end{center}
\caption{Confidence contours for the mass outflow rate versus terminal velocity parameter ($f_{\rm v}$) for the wind model, where the resultant terminal velocity is $v_{\infty}/c=0.25f_{\rm v}$ 
for the \textsc{fast32} disk wind grid. The upper panel shows the 2022 slice B epoch and the lower panel the 2017 \xmm\ epoch and the contours represent the 68\%, 90\%, 99\% and 
99.9\% significance levels for 2 interesting parameters. Both the mass outflow rate and terminal velocity are significantly higher in the 2022 slice B epoch compared to 2017. Note the different axis scale across the two plots.}
\label{fig:contour}
\end{figure}

For the wind model, the $L_{\rm X}$ parameter sets the wind ionization, scaled to the 2--10\,keV luminosity, so the $L_{\rm X}$ parameter was 
varied between the three epochs in direct proportion to the intrinsic (absorption corrected) 2--10\,keV luminosity. Relative to the brightest 2022 slice A epoch, the $L_{\rm X}$ 
parameter was set to $\times0.8$ and $\times0.45$ of this value for the slice B and the least luminous 2017 epochs respectively. 
The inclination parameter, $\mu$, was assumed to not vary between epochs, while $\dot{M}$ and the terminal velocity parameter, $f_{\rm v}$, 
were allowed to vary independently. Note that the terminal velocity for each epoch is then calculated from equation~1, where $v_{\infty}/c=0.25f_{\rm v}$ for a wind launch radius of $32R_{\rm g}$.
The input photon index of the \textsc{fast32} grid for each epoch was tied to the value determined by the power-law continuum, 
in order to best represent the slope of the ionizing continuum.   
The exception to this is for the slice A epoch (best fit value of $\Gamma=2.47\pm0.02$) 
which slightly exceeds the maximum value computed 
in the \textsc{fast32} grid. The input photon index for this epoch was subsequently fixed to $\Gamma=2.4$ in the disk wind model. 

\begin{deluxetable*}{lccc}
\tablecaption{Disk Wind Model Results to Broad-band (0.3--30\,keV) Spectra.}
\tablewidth{500pt}
\tablehead{\colhead{Parameter} & \colhead{Slice A} & \colhead{Slice B} & \colhead{XMM/2017}}
\startdata
Diskwind:-\\
$\dot{M}^{a}$ & $<0.03$ & $0.23\pm0.06$ & $0.096\pm0.011$\\
\% $L_{X}^b$ & $0.31\pm0.04$ & $0.24\pm0.03$ & $0.14\pm0.02$ \\
$f_{\rm v}^c$ & $1.05^f$ & $1.05^{+0.12}_{-0.19}$ & $0.44\pm0.03$  \\
$v_{\infty}/c^d$ & $0.26^f$ & $0.26^{+0.03}_{-0.05}$ & $0.110\pm0.008$  \\
$\mu=\cos\theta$$^e$ & $0.53\pm0.02$ & $0.53^t$ & $0.53^t$\\
\hline
Power-law:-\\
$\Gamma$ & $2.47\pm0.02$ & $2.29\pm0.09$ & $1.92\pm0.02$\\
$N_{\rm PL}^g$\,($\times10^{-3}$\,photons\,cm$^{-2}$\,s$^{-1}$\,keV$^{-1}$) & $4.4\pm0.2$ & $2.7\pm0.6$ & $0.96\pm0.06$\\
$F_{\rm 2-10\,keV}^h$\,($\times10^{-12}$\,erg\,cm$^{-2}$\,s$^{-1}$) & 4.8 & 2.3 & 1.35\\
$L_{\rm 2-10\,keV}^i$\,($\times10^{43}$\,erg\,s$^{-1}$) & $6.1\pm0.3$ & $4.8\pm1.1$ & $2.8\pm0.2$ \\ 
Constant$^j$ & $0.86\pm0.02$ & & $1.05\pm0.01$\\
\hline
Comptt:-\\
$kT ({\rm eV})$ & $180\pm7$ & & $139\pm6$ \\
$\tau$ & $19.0\pm0.9$ & & $21.5^{+1.1}_{-1.6}$ \\
$F_{\rm 0.3-2\,keV}^k$\,($\times10^{-12}$\,erg\,cm$^{-2}$\,s$^{-1}$) & 13.9 & & 3.1 \\
$L_{\rm 0.3-2\,keV}^l$\,($\times10^{44}$\,erg\,s$^{-1}$) & $2.7\pm0.2$ & & $1.1\pm0.1$\\
\hline
Gaussian (Fe K$\alpha$):-\\
$E (\rm {keV})$ & $6.24^{+0.18}_{-0.21}$ & $6.24^f$ & $6.24^f$\\
$\sigma (\rm{keV})$ & $0.84^{+0.27}_{-0.21}$ & $0.84^f$ & $0.84^f$\\
$N_{\rm gauss}^m$\,($\times10^{44}$\,erg\,s$^{-1}$) & $2.1^{+0.6}_{-0.5}$ & $<1.5$ & $<0.6$\\
${\rm EW} (\rm{eV})$ & $360\pm100$ & $<350$ & $<170$
\enddata
\tablenotetext{a}{Mass outflow rate in Eddington units.}
\tablenotetext{b}{Percentage ionizing ($2-10$\,keV) luminosity to Eddington luminosity.}
\tablenotetext{c}{Terminal velocity parameter, $f_{\rm v}$ as defined in equation 1.}
\tablenotetext{d}{Wind terminal velocity}
\tablenotetext{e}{Cosine of wind inclination, wrt the Polar axis.}
\tablenotetext{f}{Parameter is fixed.}
\tablenotetext{g}{Power-law normalization at 1\,keV.}
\tablenotetext{h}{Observed 2--10\,keV flux.}
\tablenotetext{i}{Intrinsic 2--10\,keV luminosity, corrected for absorption.}
\tablenotetext{j}{Multiplicative cross normalization constant between EPIC-pn and NuSTAR in 2022 and EPIC-MOS and pn respectively in 2017 and 2022.}
\tablenotetext{k}{Observed 0.3--2\,keV flux.}
\tablenotetext{l}{Intrinsic 0.3--2\,keV luminosity, corrected for absorption.}
\tablenotetext{m}{Photon flux of Fe K$\alpha$ emission line, after including the disk wind component.}
\tablenotetext{t}{Denotes parameter is tied between observations.}
\label{tab:wind}
\end{deluxetable*}

The results of the diskwind fits are shown in Table~\ref{tab:wind}, while the fit to the broad-band spectra is displayed in Figure~\ref{fig:wind}. Overall the best-fit inclination parameter is 
$\mu=0.53\pm0.02$, corresponding to an inclination angle of $\theta=58\pm2\degg$, placing our line of sight inside of the wind opening angle of 45 degrees. 
The fit statistic to all of the spectra is acceptable
($\chi^2/\nu=1758.0/1635$) and provides a significant improvement upon the baseline model without including the disk wind component, where $\chi^2/\nu=2003.6/1642$, which 
is rejected with a null hypothesis probability of $N=1.7\times10^{-9}$. There are no strong residuals present against the best fit wind model (Figure~\ref{fig:wind}, lower panel), while in the {\it NuSTAR} slice B and 2017 epochs, the disk wind model is able to account for the absorption profiles in the iron K band. 

While the disk wind model is able to simultaneously account for the Fe K emission and absorption in the slice B and 2017 epochs, it does not account for the broad Fe K$\alpha$ emission present in the slice A epoch (see Table~\ref{tab:wind} for parameters). 
This emission may originate from an ionized disk reflector \citep{Garcia14}, which is often observed in the X-ray spectra of bare Seyfert 1s (e.g. \citealt{Porquet18, Porquet21}). 
Alternatively, it  could arise from additional scattering of wind photons from gas out of the line of sight. 
While there is no requirement for any disk wind absorption in the bright slice A epoch, the additional Fe K$\alpha$ emission  
might hint at some asymmetry or inhomogeneity to the wind structure. Thus while our line of sight is devoid of absorbing gas (e.g. lower column or covering), the additional iron K$\alpha$ emission could still arise 
via scattering from denser gas out of line of sight, which may not have varied over time.


As for the disk wind model parameters, a strong variation in mass outflow rate is required between the three epochs, varying (in Eddington units) from $\dot{M}<0.03$ (slice A) 
to $\dot{M}=0.23\pm0.06$ (slice B) to $\dot{M}=0.096\pm0.011$ (2017). Note that the upper-limit upon $\dot{M}$ for slice A is obtained upon assuming the same wind velocity as per slice B. 
Given the strong iron K emission and absorption seen towards slice B, its mass outflow rate has the highest value of all the epochs and is also about a factor of two higher than in the 2017 spectrum. The disk wind analysis also confirms the variation in terminal wind velocity between the slice B interval ($v_{\infty}/c=0.26^{+0.03}_{-0.05}$) and of the 2017 spectrum (where $v_{\infty}/c=0.110\pm0.008$). 

Figure~\ref{fig:contour} also shows confidence contours between the mass outflow rate and terminal velocity parameter ($f_{\rm v}$) for the slice B and 2017 epochs. 
The difference between the mass outflow rate for these epochs largely result from the higher terminal velocity of the slice B epoch, which may be expected given that $\dot{M} \propto v_{\infty}$. In the slice B contours, there is a hint of a second solution 
at an even higher velocity ($f_{\rm v}\sim1.4$), although this is not formally required by the data. The extension of the contours in velocity space for slice B likely results from the breadth of the profile 
towards higher energies in the {\it NuSTAR} spectrum, while in the 2017 epoch, the velocity width of the profile is smaller, the contours are symmetrical and the velocity space is highly constrained. 

It is also worth noting the variation in intrinsic luminosity between the 3 epochs. After correcting for the intrinsic absorption imparted by the wind, the 2--10\,keV luminosity for the slice B epoch is inferred 
to be within 80\% of the value for the bright Slice A epoch. This means that changes in the $L_{\rm X}$ parameter alone are not sufficient to account for the opacity change between slices A and B and are instead driven by variations in mass outflow rate. This implies that the wind is intrinsically variable even down to short timescales of tens of kilo-seconds. The implications will be discussed in Section~7.

The increase in $\dot{M}$ from slice A to B has two effects, 
it increases the line of sight column through the wind and consequently, it also decreases the ionization, as a result of the wind density being higher. 
Indeed for slice B, the continuum is suppressed by a factor of two by the wind, primarily through electron scattering, as well as via some bound--free and bound--bound absorption at iron K.  As electron scattering decreases the continuum by a factor of $e^{-N_{\rm H}\sigma_{T}}$ (where $\sigma_{\rm T}$ is the Thomson cross-section), this implies that the column density along the wind is approximately $N_{\rm H}\sim10^{24}$\,cm$^{-2}$, in broad agreement with the \textsc{xstar} results in Section~4.  One interesting consequence of this 
is that the factor of $\times2$ flux drop observed in slice B may be accounted for by this increase in column. In contrast, the correction for intrinsic wind absorption alone is not sufficient to account for the lower flux observed towards the 2017 epoch. 

\section{The Soft X-ray Disk Wind}

The 2017 and 2022 \xmm\ RGS spectra were also analyzed in order to quantify any absorption from both the fast disk wind and from any lower velocity gas from a soft X-ray warm absorber. 
In the \pg\ spectra, the RGS bandpass is sensitive over the range from 0.45--1.9\,keV.  
The same baseline model as per equation~2 was used, with a warm absorber included (modeled by \textsc{xstar}) to also assess the 
presence of lower ionization, low velocity gas, as is frequently observed in many Seyfert 1 galaxies (e.g., \citealt{Kaastra00, Kaspi02, Crenshaw03}). 
Additional emission lines, in this case likely from the He-like lines of O\,\textsc{vii} and Ne\,\textsc{ix}, are also included. The 
warm absorber and emission are described in more detail in Section~6.1. 
For the continuum, the soft X-ray excess, in the form of the Comptonized disk emission (\textsc{comptt} component) dominates 
the soft band below 1\,keV (or wavelengths longer than 12\,\AA). The power-law emission is required to fit the short wavelength tail 
of the spectra, where the photon indices of the 2017 and 2022 spectra have been fixed at $\Gamma=1.9$ and $\Gamma=2.4$ respectively, 
as per the broad-band analysis, as otherwise they are not well determined. 

\begin{figure}
\begin{center}
\rotatebox{-90}{\includegraphics[height=8cm]{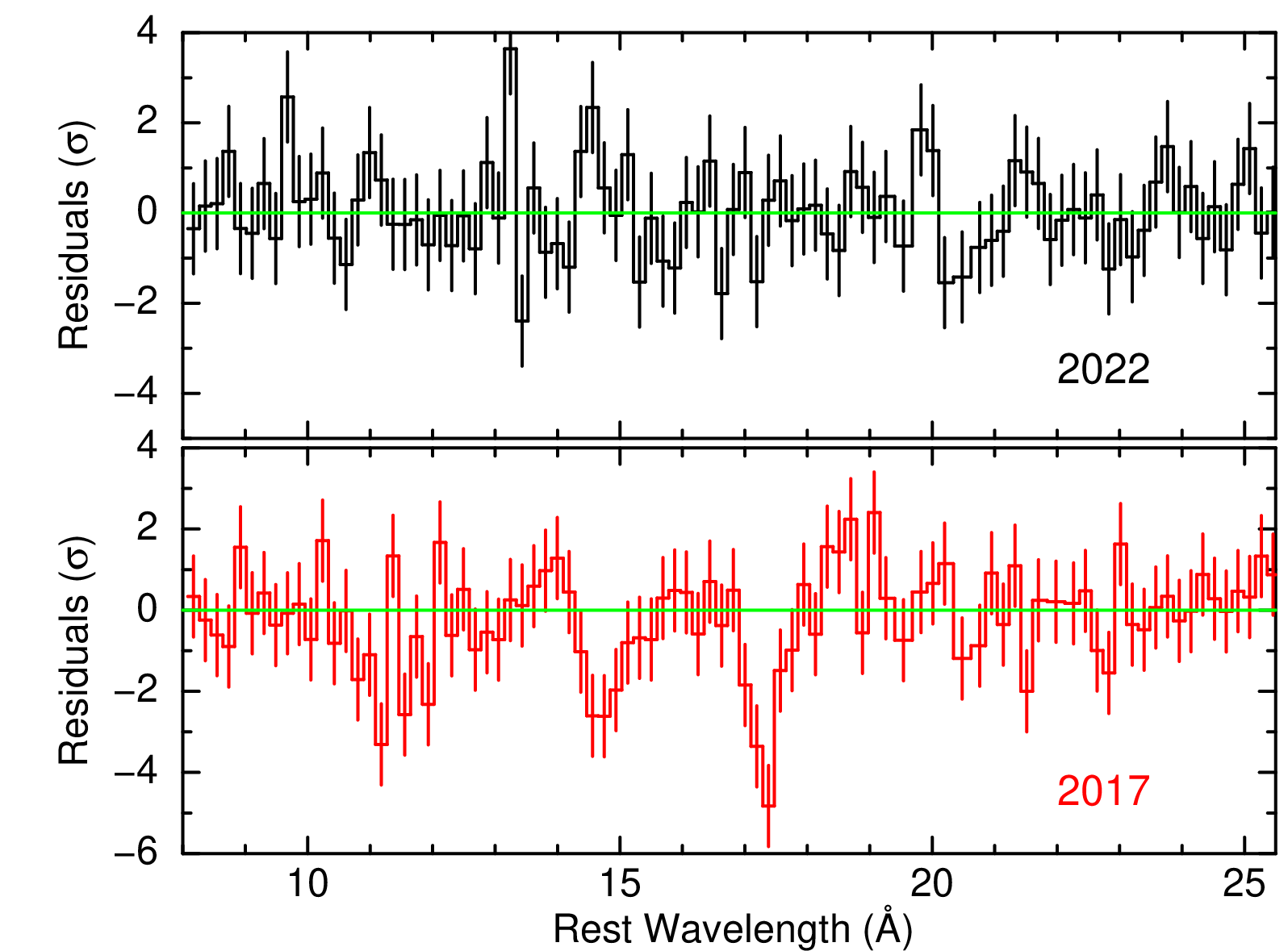}}
\end{center}
\caption{Residuals to the soft X-ray \xmm\ RGS spectra of \pg, with respect to the baseline model without a disk wind absorber. The upper panel (black points) shows the 
high flux 2022 spectrum and the lower panel (red points) shows the low flux 2017 spectrum. Note that the soft X-ray flux has varied by a factor of $\times3$ at 17.3\,\AA\ between the observations, see Table~5. 
In the 2022 spectrum, no significant residuals are present against the baseline model, which accounts for an additional warm absorber seen towards \pg. However the 2017 RGS spectrum 
shows a highly significant ($5\sigma$) broad absorption trough at 17.3\,\AA, which may be associated to the O\,\textsc{viii} Ly$\alpha$ line which is blue-shifted by $0.1c$. 
A second, weaker trough at 14.7\,\AA\ may arise from a higher velocity component of the wind.}
\label{fig:rgs}
\end{figure}

The residuals of each of the RGS spectra to the baseline model without the disk wind absorber are shown in Figure~\ref{fig:rgs}. The baseline model provides a good fit to the 2022 spectrum and 
no significant residuals remain (see upper panel). In contrast, the 2017 residuals show a strong absorption line at the $5\sigma$ level, measured at 
$\lambda=17.32\pm0.10$\,\AA\ (or $E=0.716\pm0.003$\,keV) in the AGN rest frame, while a second weaker feature (at the $3\sigma$ level) may also be present 
at $\lambda=14.76\pm0.15$\,\AA\ (or $E=0.840\pm0.009$\,keV). In terms of their significance when fitted with a Gaussian absorption line, the first feature improves the fit by $\Delta\chi^2=36.0$ (for $\Delta\nu=3$ 
fewer degrees of freedom), while for the second feature $\Delta\chi^2=15.9$ for $\Delta\nu=3$. 
Considering the identification of the 17.32\,\AA\ line, there is no obvious atomic line present 
at this wavelength in the rest frame; the closest line is the higher order $1s\rightarrow5p$ O\,\textsc{vii} transition, which would be expected to be weak. However it may be identified with 
a strong blue-shifted resonance line from O\,\textsc{viii} Ly$\alpha$ ($1s\rightarrow2p$) which occurs at a rest wavelength of 18.97\,\AA; in this case the inferred outflow velocity would be $v/c\approx0.09$, which is very similar 
to the velocity inferred from the blueshifted iron K absorption at 7.5\,keV during this epoch. Then in the case of the shorter wavelength trough, it could either arise from the weaker O\,\textsc{viii} Ly$\beta$ 
line at 16\,\AA\ (e.g. see \citealt{Sidoli01}) at a similar velocity or from an even higher velocity component of the O\,\textsc{viii} Ly$\alpha$ line.

\begin{figure*}
\begin{center}
\rotatebox{-90}{\includegraphics[height=8cm]{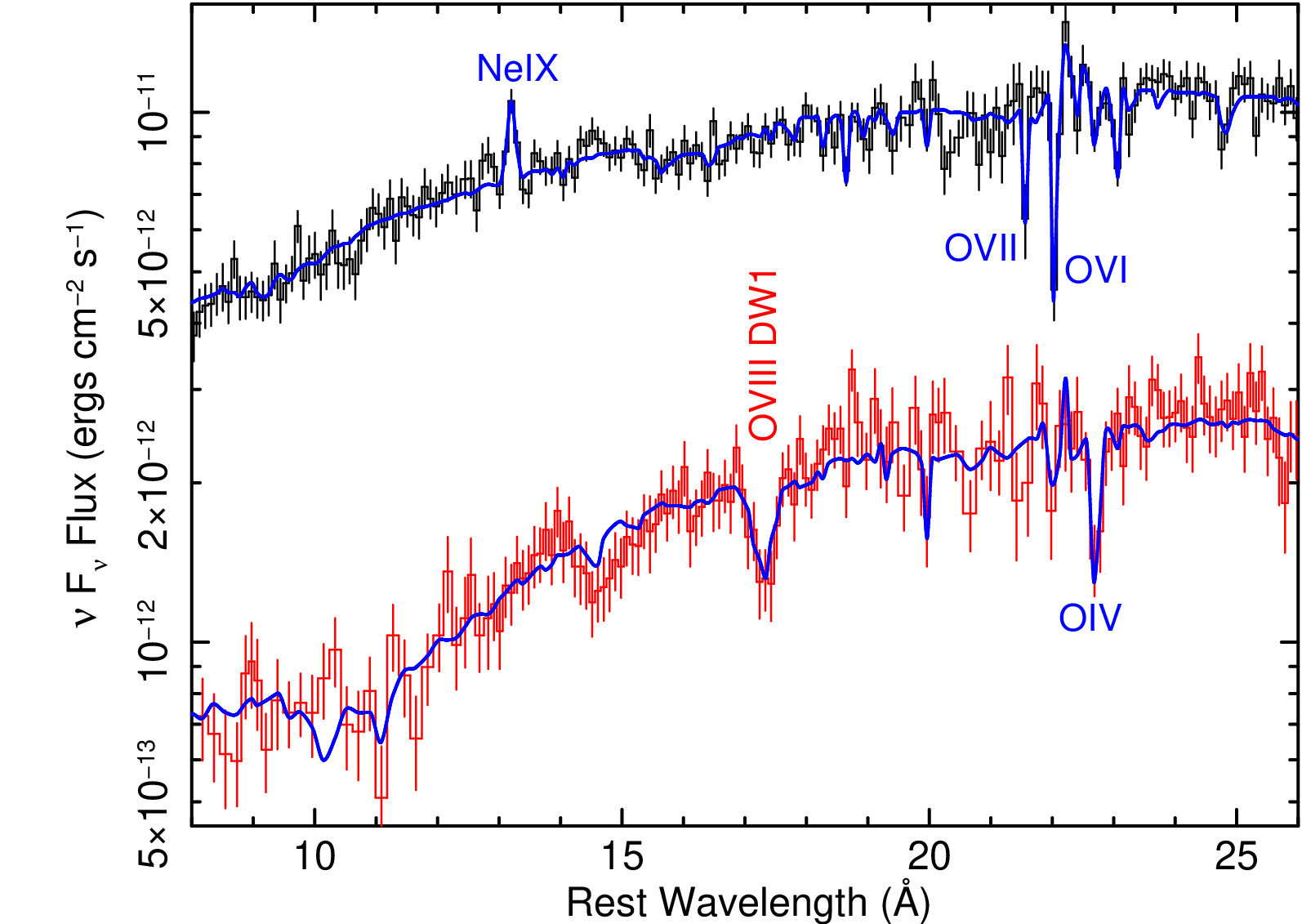}}
\rotatebox{-90}{\includegraphics[height=8cm]{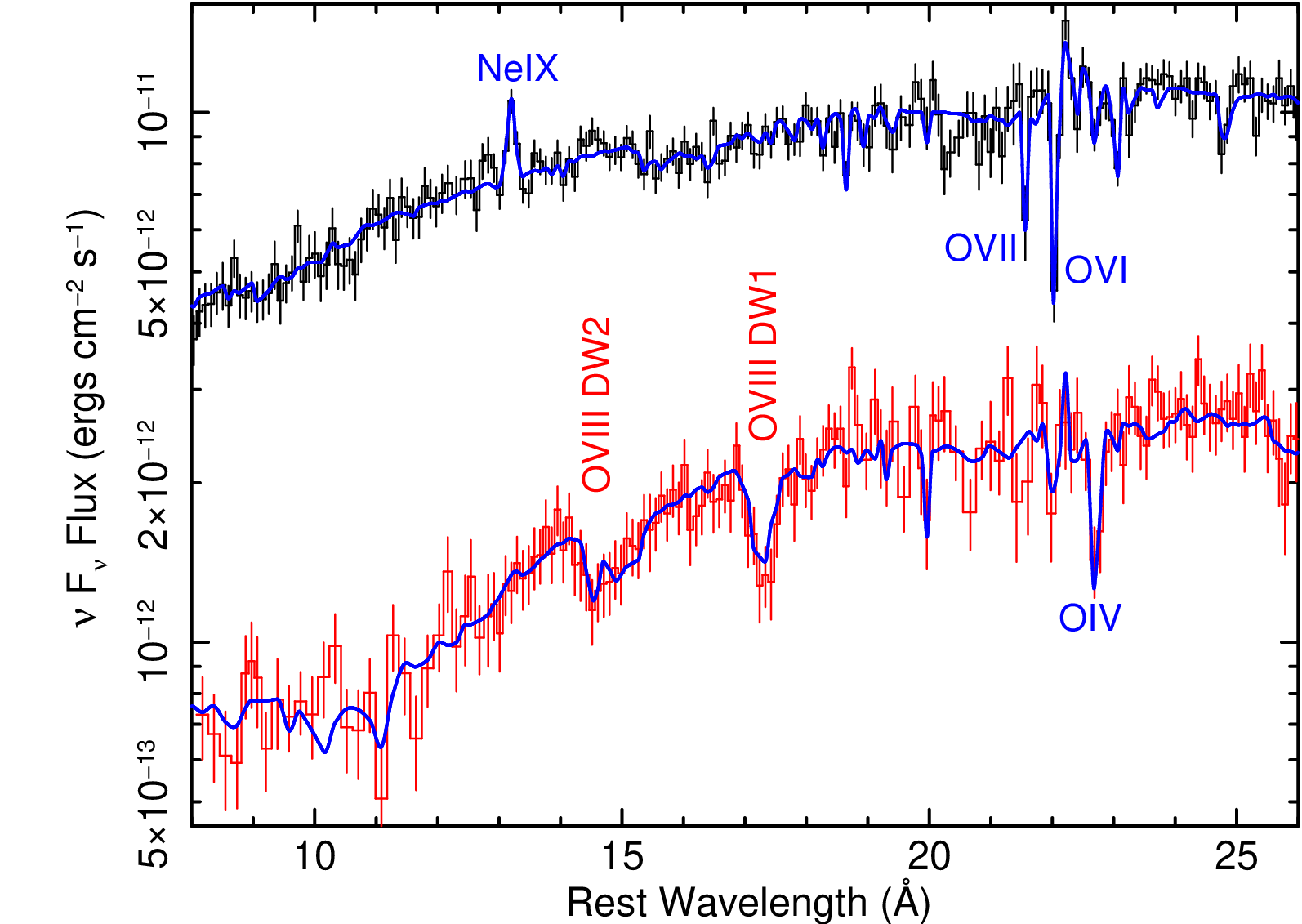}}
\end{center}
\caption{Fluxed RGS spectra of \pg, for the 2017 (red) and 2022 (black) observations. The solid blue line shows the best-fit model to both spectra. The left panel shows the spectra fitted with a single disk wind absorption zone (red label, O\,\textsc{viii} DW1), with a terminal velocity of $0.1c$, which can account for the strong 17.3\,\AA\ absorption trough. The velocity of this absorber is also consistent with what is obtained from the iron K absorption profile in 2017. In contrast no signature of the disk wind is present in the high flux 2022 spectrum. 
The right panel shows the addition of a second, faster ($0.3c$) disk wind zone to the 2017 spectrum (red label, O\,\textsc{viii} DW2), which can model the second absorption trough, but it is of lower significance. Note the most prominent absorption and emission lines from a low velocity warm absorber are also marked in blue. }
\label{fig:rgswind}
\end{figure*}

As per Section~5, the disk wind model is then applied to the RGS spectra. Following the broad-band fit, the $L_{\rm X}$ parameter between 2017 and 2022 is tied in proportion to the inferred intrinsic 2--10\,keV 
luminosity; in this case the 2017 epoch is a factor of $0.45\times$ the luminosity of the bright 2022 epoch. The spectral fitting results are listed in Table~\ref{tab:rgswind} and a fit to the RGS spectra using a single disk wind zone is shown in Figure~\ref{fig:rgswind} (left panel). Overall the fit statistic is good with $\chi^2/\nu=572.5/545$, while the fit without including the disk wind is significantly worse with $\chi^2/\nu=628.5/550$ (and $\Delta\chi^2=56$ for $\Delta\nu=5$). The wind model is able to reproduce the absorption feature near 17.3\AA\ in the 2017 spectrum (shown in red) in the form of a blue-shifted O\,\textsc{viii} Ly$\alpha$ line. The velocity of the disk wind component is $v_{\infty}/c=-0.098\pm0.010$, 
which is in agreement with the wind velocity inferred from the iron K absorption profile in the previous section. 
It is also in agreement with the analysis presented by \citet{Kosec20}, who reported a wind velocity of $-26900\pm600$\,km\,s$^{-1}$ from their analysis of the same RGS spectrum. 

The 2017 mass outflow rate, of $\dot{M}=0.13\pm0.03$, is also in broad agreement with the iron K zone, while the inclination is also consistent (see Table~\ref{tab:rgswind}). In contrast, no disk wind component is required 
in the bright state 2022 epoch, with an upper-limit to the mass outflow rate of $\dot{M}<0.05$ assuming the same terminal velocity. This is not surprising, as the 2022 \xmm\ RGS spectrum coincided with the 
bright and featureless slice A portion of the {\it NuSTAR} observation where no iron K wind features were present either. 
Alternatively, if we assume that the mass outflow rate remained constant between the 
2017 and 2022 spectra, then the ionizing luminosity of the former epoch has to be much lower than the latter, with $L_{\rm X}=0.14\pm0.03$ versus $L_{\rm X}>0.9$ respectively, in order to 
explain the difference in opacity. 

The second absorption trough at 14.7\,\AA\ is not fully accounted for by the disk wind model in the 2017 spectrum, where the model imparts only a weak absorption feature due to 
O\,\textsc{viii} Ly$\beta$ at the above terminal velocity as in 2017. However, this second trough can be modeled with a faster disk wind component originating from 
O\,\textsc{viii} Ly$\alpha$ (see Figure~\ref{fig:rgswind}, right panel). In this case, the corresponding outflow velocity is $v_{\infty}/c=-0.31\pm0.02$ (or $f_{\rm v}=1.25\pm0.09$), with a corresponding 
mass outflow rate of $\dot{M}=0.23^{+0.10}_{-0.06}$. Interestingly, both the velocity and outflow rate are consistent with what is found during the slice B interval of the {\it NuSTAR} 
spectrum. Note that the improvement in fit statistic for this second faster zone ($\Delta\chi^2=16$ for $\Delta\nu=3$) is much smaller than for the above $0.1c$ wind zone. 
Furthermore, as the 2017 \xmm\  observation was not performed simultaneously with {\it NuSTAR}, it is not possible to verify whether this faster component is present in the iron K 
band above 9\,keV. 

\begin{deluxetable}{lcc}
\tablecaption{Model Results to soft X-ray (0.4--2.0\,kev) RGS Spectra.}
\tablewidth{250pt}
\tablehead{\colhead{Parameter} & \colhead{2017} & \colhead{2022}}
\startdata
Diskwind:-\\
$\dot{M}^{a}$ & $0.13\pm0.03$ & $<0.05$ \\
\% $L_{X}^b$ & $0.14\pm0.03$ & $0.32\pm0.07$ \\
$f_{\rm v}^c$ & $0.39\pm0.04$ & $0.39^t$  \\
$v_{\infty}/c^d$ & $0.098\pm0.010$ & $0.098^t$ \\
$\mu=\cos\theta$$^e$ & $0.54\pm0.02$ & $0.54^t$ \\
\hline
Power-law:-\\
$\Gamma$ & $1.9^f$ & $2.4^f$ \\
$N_{\rm PL}^g$\,($\times10^{-3}$\,photons\,cm$^{-2}$\,s$^{-1}$\,keV$^{-1}$) & $1.4\pm0.3$ & $3.8\pm0.2$ \\
\hline
Comptt:-\\
$kT ({\rm eV})$ & $113^{+11}_{-8}$ & $171^{+17}_{-14}$ \\
$\tau$ & $>35$ & $22.5^{+3.5}_{-2.8}$ \\
$F_{\rm 0.3-2\,keV}^h$\,($\times10^{-12}$\,erg\,cm$^{-2}$\,s$^{-1}$) & 2.8 & 13.7 \\
$L_{\rm 0.3-2\,keV}^{i}$\,($\times10^{-12}$\,erg\,cm$^{-2}$\,s$^{-1}$) & $1.1\pm0.1$ & $2.7\pm0.2$\\
\hline
Warm Abs:-\\
$N_{\rm H}$\,($\times10^{20}$\,cm$^{-2}$) & $3.8^{+1.5}_{-1.3}$ & $3.8^t$ \\
$\log(\xi/{\rm erg\,cm\,s}^{-1})$ & $-0.47\pm0.23$ & $0.69\pm0.30$\\
$v_{\rm out}^j$\,(km\,s$^{-1}$) & $<360^t$ & $<360$\\
$v_{\rm turb}^j$\,(km\,s$^{-1}$) & $200^f$ & $200^f$ \\
\hline
Gaussian (O\,\textsc{vii}):-\\
$E (\rm{eV})$ & $556^t$ & $556\pm2$ \\
$\sigma (\rm{eV})$ & $<6^t$& $<6$ \\
Line Flux\,($\times10^{-5}$\,photons\,cm$^{-2}$\,s$^{-1}$) & $<1.8$ & $4.9\pm2.3$ \\
${\rm EW} (\rm{eV})$ & $<1.5$ & $2.1\pm1.0$ \\
\hline
Gaussian (Ne\,\textsc{ix}):-\\
$E (\rm{eV})$ & $939^t$ & $930\pm2$ \\
$\sigma (\rm{eV})$ & $<4.2^t$& $4.2\pm2.1$ \\
Line Flux\,($\times10^{-5}$\,photons\,cm$^{-2}$\,s$^{-1}$) & $<0.4$ & $3.0\pm0.9$ \\
${\rm EW} (\rm{eV})$ & $<3.6$ & $4.7\pm1.5$ 
\enddata
\tablenotetext{a}{Mass outflow rate in Eddington units.}
\tablenotetext{b}{Percentage ionizing ($2-10$\,keV) luminosity to Eddington luminosity.}
\tablenotetext{c}{Terminal velocity parameter, $f_{\rm v}$ as defined in equation 1.}
\tablenotetext{d}{Wind terminal velocity}
\tablenotetext{e}{Cosine of wind inclination, wrt the Polar axis.}
\tablenotetext{f}{Parameter is fixed.}
\tablenotetext{g}{Power-law normalization at 1\,keV.}
\tablenotetext{h}{Observed 0.3--2\,keV flux.}
\tablenotetext{i}{Intrinsic 0.3--2\,keV luminosity, corrected for absorption.}
\tablenotetext{j}{Outflow and turbulence velocity of the \textsc{xstar} warm absorber, in units of km\,s$^{-1}$.}
\tablenotetext{t}{Denotes parameter is tied between observations.}
\label{tab:rgswind}
\end{deluxetable}

\subsection{Properties of the soft warm absorption and emission}

\begin{figure}
\begin{center}
\rotatebox{-90}{\includegraphics[height=8cm]{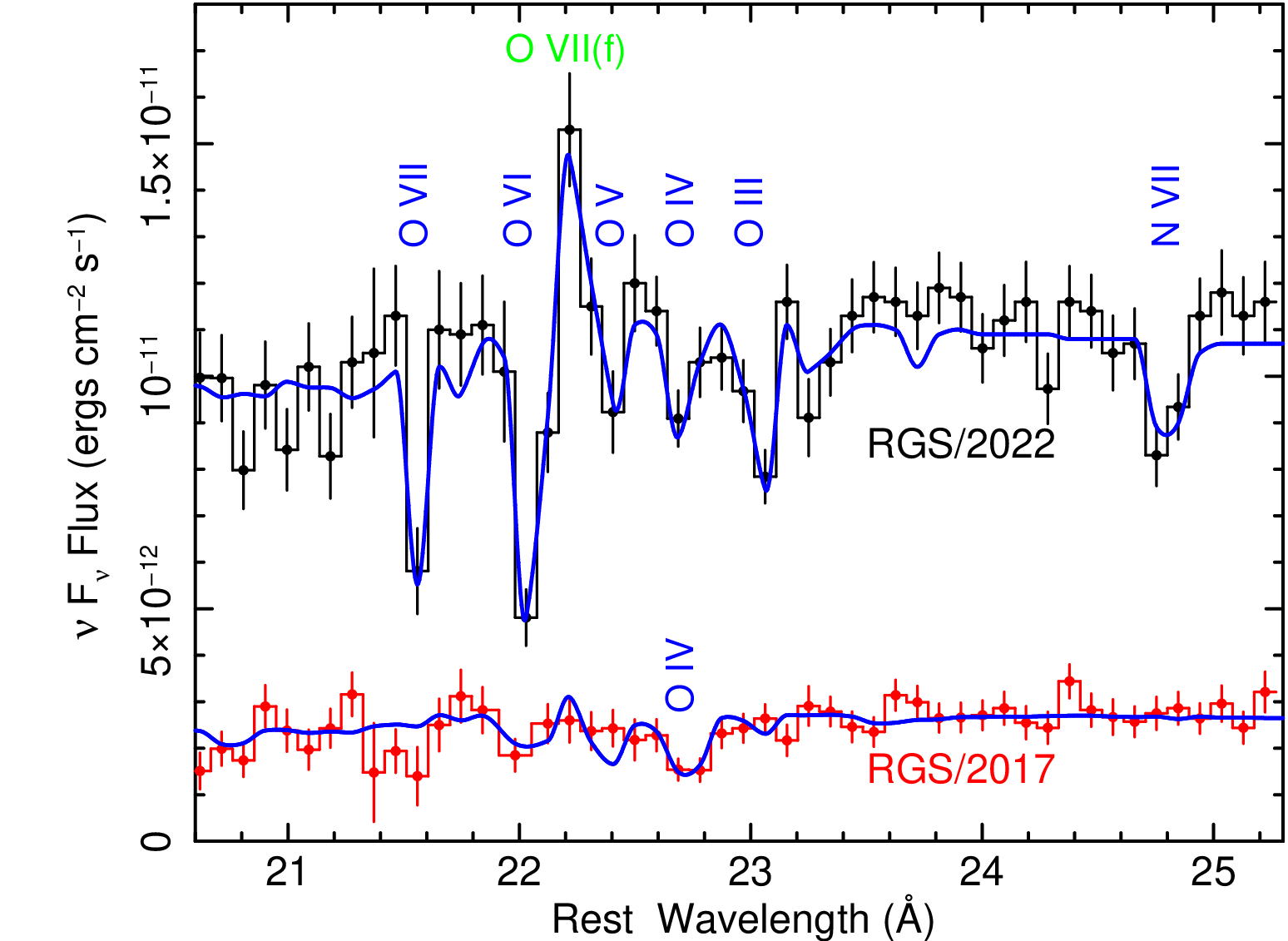}}
\end{center}
\caption{Zoom into the O K-shell region for both the 2017 and 2022 RGS spectra. The absorption lines originating from a low velocity warm absorber (lines marked with blue labels) are strongest in the high flux 
2022 observation. In comparison, the highest excitation lines due to O\,\textsc{vi-vii} and N\,\textsc{vii} are absent in 2017. This might suggest the warm absorber is under-ionized in the 2017 epoch. 
Similarly, emission due to the O\,\textsc{vii} forbidden line (marked with a green label) is present in 2022, but not in 2017.}
\label{fig:oxygen}
\end{figure}

In addition to the fast disk wind, a soft X-ray warm absorber is also present in \pg. The main features of the warm absorber and emitter are marked with blue labels on the RGS spectra in Figure~\ref{fig:rgswind}, while the properties of the warm absorber and two possible emission lines (most likely from O\,\textsc{vii} and Ne\,\textsc{ix}) are listed in Table~\ref{tab:rgswind}. The largest imprint of the warm absorber occurs in the Oxygen K-shell band, which is illustrated in the zoom into this region in Figure~\ref{fig:oxygen}. The warm absorption lines are most prominent against the high flux 2022 epoch. The strongest absorption in 2022 is due to the 
higher excitation lines of O\,\textsc{vi}, O\,\textsc{vii} and N\,\textsc{vii}, with weaker absorption from lower ionization gas (O\,\textsc{iii-v}). Conversely, the higher excitation lines are not present in the 
low flux 2017 spectrum and only the low ionization absorption (e.g. O\,\textsc{iv}) is present. From fitting the two epochs simultaneously, the change in absorption opacity can be explained by 
an increase in absorber ionization from 2017 to 2022. In the \textsc{xstar} fits listed in Table~\ref{tab:rgswind}, this is accounted for by a simple change in the ionization parameter from $\log\xi=-0.47\pm0.24$ (2017) to 
$\log\xi=0.69\pm0.30$ and assuming a constant column density of $N_{\rm H}=3.8^{+1.5}_{-1.3}\times10^{20}$\,cm$^{-2}$. In this simple scenario, the 2017 absorber is de-ionized with respect to the 2022 
absorber, resulting from the $\times5$ lower observed soft X-ray flux. Unlike the fast wind, the warm absorber does not require any velocity shift, where the upper-limit on its outflow velocity is $<360$\,km\,s$^{-1}$. The column density is also at least 3 orders of magnitude below that of the fast wind, as was derived in Section~4.

Two emission lines are required in the bright state 2022 spectrum. The first emission line occurs at $\lambda=22.3\pm0.1$\,\AA\ (or $556\pm2$\,eV), which is close to the expected position 
of the O\,\textsc{vii} forbidden line (at 22.1\,\AA\ or 561\,eV). The slight redshift may just be as a result of the strong O\,\textsc{vi} absorption line on the blue-wards side of the emission (see Figure~\ref{fig:oxygen}). 
As a result of this, only an upper-limit is placed on the line width, of $\sigma<6$\,eV, corresponding to $\sigma_{\rm v}<3200$\,km\,s$^{-1}$, which would be consistent with either a BLR or NLR origin. 
In comparison, only an upper-limit is placed on the O\,\textsc{vii} line flux in the low flux 2017 spectrum (see Table~\ref{tab:rgswind}), which might suggest it responds to the continuum flux, although the five years between observations prohibits a meaningful constraint on its timescale. A second emission line in 2022 is seen at $939\pm2$\,eV (or $13.2\pm0.1$\,\AA, see Figure~\ref{fig:rgswind}), which is somewhat blue-shifted with respect to the expected position of the Ne\,\textsc{ix} triplet emission (between 905--921\,eV or 13.4--13.7\,\AA). Given its blue-shift, its origin is more uncertain, although its addition to the model is highly significant ($\Delta\chi^2=30.9$). No other emission lines are detected in the soft X-ray band. 

\section{Discussion} \label{sec:discussion}

{\it NuSTAR} and \xmm\ observations of \pg\ in 2022 reveal the remarkable variability of its ultra fast outflow. On short timescales, the wind opacity appears to increase by at least a factor of three during the last 60\,ks of the {\it NuSTAR} observation (the slice B spectrum), which corresponds to an interval in the lightcurve where the X-ray flux of \pg\ decreased (Section~3). This flux drop interval revealed a broad iron K absorption trough centered near 9 keV and implies an outflow velocity of $\sim0.3c$ (Section~4). The onset of the flux drop occurred on very short timescales, of $\sim10$\,ks, corresponding to 2 {\it NuSTAR} satellite orbits, highlighting the rapid wind variability.  
On longer timescales, between the 2017 and 2022 epochs, a substantial change in the terminal wind velocity was 
observed, as seen by from the blueshift of the iron K absorption profile, varying from $v/c=-0.10\pm0.01$ in 2017 to $v/c=-0.26\pm0.04$ in 2022, when modeled with a physically realistic disk wind model (Section~5). Changes in the opacity of the soft X-ray wind is also observed between the 2017 and 2022 epochs (Section~6), where the 2017 low flux RGS spectrum showed a deep absorption trough due to a broad O\,\textsc{viii} Ly$\alpha$ line and which has diminished in the 
$\times5$ higher flux 2022 RGS spectrum.

\pg\ thus appears to exhibit all of the variability characteristics seen towards other AGN disk winds. For instance, in the highly variable NLS1 IRAS 13224$-$3809, the fast wind at iron K and at soft X-rays 
appears to respond to the X-ray flux of the AGN, becoming transparent at high fluxes, potentially due to the increase in wind ionization \citep{Parker17, Pinto18}. The same effect occurs in PDS\,456, 
where the X-ray spectrum can become almost featureless at very high flux, corresponding to flaring states, compared to the more quiescent X-ray flux states  where the strong wind absorption profiles are more commonly observed \citep{Reeves21}. 
A similar behavior occurs in \pg\ comparing the 2022 high flux slice A spectrum, which is bare and devoid of wind features, with the low flux 2017 X-ray spectrum (e.g., \citealt{Kosec20, Laurenti21}). 
The increase in opacity in the last 60\,ks of the 2022 {\it NuSTAR} observation of \pg\ may arise due to a rapid wind obscuration event, with an increase in column density of $\Delta N_{\rm H}\approx10^{24}$\,cm$^{-2}$ 
or by an equivalent change in covering fraction.
Similar opacity variations have also been observed in other ultra fast outflows on timescales down to a day; e.g. \mcg\ \citep{Braito18}, PG\,1211+143 \citep{ReevesLobbanPounds18}, PDS\,456 \citep{Gofford14, Matzeu16, Reeves18}. These have been interpreted as being due to clumpy, inhomogeneous winds, where opacity variations are observed over size-scales of tens of gravitational radii. 

Finally, \pg\ appears to exhibit a remarkable change in its terminal velocity, by as much as a factor of $\times3$ between 2017 and 2022. Variations in the outflow velocities of fast outflows have been established in several 
cases, most notably in PDS\,456 \citep{Matzeu17}, APM 08279$+$5255 \citep{SaezChartas11}, IRAS\,13224$-$3809 \citep{ChartasCanas18} and \mcg\ \citep{Braito21, Braito22}. 
As we discuss further below, the example which may bare the closest behavior to \pg\ is in \mcg, where a similar factor of $\times3$ variability in wind velocity is also observed, on timescales as short as 2 weeks.

\subsection{Kinematics of the X-ray wind}

The observed increase in wind velocity between the 2017 and 2022 epochs in \pg\ likely has implications for the resulting mechanical power of the disk wind. To compare the wind energetics for the different observational epochs 
(2022 slice A, 2022 slice B and 2017), the mass outflow rates and terminal velocities derived from the disk wind modeling in Section~5 are adopted, where the mass outflow rate ($\dot{M}$) is expressed in 
Eddington units. Here, the disk wind model assumes a bi-conical axisymmetric geometry, for a wind opening of 45 degrees, as is described in Section\,5. 
For comparison, the soft X-ray wind energetics derived from the 2017 epoch are also computed for comparison (from Section~6). The kinetic power in Eddington units ($\dot{E}$) is subsequently:-

\begin{equation}
\dot{E}=\frac{L_{\rm kin}}{L_{\rm Edd}} = \frac{1}{\eta} (\gamma -1) \dot{M} \approx \frac{1}{2\eta} \dot{M} \left(\frac{v}{c}\right)^2
\end{equation}

\noindent where $\gamma$ is the Lorentz factor and $\eta$ is the accretion efficiency and the equation tends to the non-relativistic limit on the RHS. 
even for the fastest slice B interval. The corresponding wind momentum thrust in Eddington units is subsequently:-

\begin{equation}
\dot{p}=\frac{\dot{p_{\rm out}}}{\dot{p}_{\rm Edd}} = \frac{1}{\eta} \dot{M} \frac{v}{c}.
\end{equation}

The disk wind energetics of \pg\ are listed in Table~\ref{tab:energetics}. We do not compute the 2022 slice A epoch, due to the lack of any measurable line of sight absorption at that time.  
However, comparing the epochs with pronounced wind absorption, there is a strong increase in the wind energetics comparing the 2017 and 2022 (slice B) epochs. 
This is largely driven by the increase in wind terminal speed from $0.11c$ to $0.26c$ between 2017 and 2022, 
as the parameters scale with velocity as $\dot{M} \propto v$ and thus subsequently $\dot{p} \propto v^2$ and $\dot{E} \propto v^3$.  As a result, 
the derived kinetic power increases by an order of magnitude from $0.6\pm0.1$\% of Eddington in 2017 (seen both in the EPIC and RGS spectra) to $8.2\pm2.5$\% during the fast wind epoch in 2022. 
Such a large variation in velocity and kinetic power is similar to what has been previously discovered in the fast wind in the Seyfert\,2 galaxy, \mcg. In this AGN, \citet{Braito22} demonstrated that the terminal velocity varied from 
$0.07c$ up to $0.2c$ in a timescale of only 2 weeks and consequently by an order of magnitude in the wind kinetic power. In contrast, in the well studied wind in PDS\,456, the 
velocity changes are more modest, ranging from $0.25-0.35c$, where \citep{Matzeu17} showed that the wind speed scaled with the X-ray luminosity as $v \propto L_{\rm X}^{0.25}$. 
As there are only 2 epochs where the wind is detected in \pg, it is not yet possible to ascertain how its speed and power scales with luminosity, e.g. whether the increase in wind velocity is a result of increased radiation pressure 
or magnetic driving \citep{Fukumura18}. 

\begin{deluxetable}{lccc}
\tablecaption{Derived Outflow Energetics for \pg.}
\tablewidth{250pt}
\tablehead{\colhead{Value} & \colhead{2022/Slice B} & \colhead{2017/EPIC} & \colhead{2017/RGS}}
\startdata
$v_{\infty}/c$ & $0.26\pm0.04$ & $0.110\pm0.008$ & $0.10\pm0.01$ \\
$\dot{M}^{a}$ & $0.23\pm0.05$ & $0.096\pm0.011$ & $0.13\pm0.02$\\
$\dot{E}^b$\,\% & $8.2\pm2.5$ & $0.58\pm0.09$ & $0.65\pm0.13$\\
$\dot{p}^{c}$  & $0.62\pm0.16$ & $0.11\pm0.02$ & $0.13\pm0.02$ 
\enddata
\tablenotetext{a}{Mass outflow rate in Eddington units, as per Tables 4 and 5.} 
\tablenotetext{b}{Outflow kinetic power as a percentage of the Eddington luminosity.}
\tablenotetext{c}{Outflow momentum rate in Eddington units.}
\label{tab:energetics}
\end{deluxetable}

\subsection{Short Timescale Wind Variability}

The wind in \pg\ is variable on short timescales, as evidenced by the change in opacity in the 9\,keV absorption trough between the slice A and slice B intervals in the 2022 {\it NuSTAR} observation. 
As was shown in Section~3, this coincides with the dip in the {\it NuSTAR} lightcurve, whose onset occurs within a timescale of $\Delta t=10$\,ks and may correspond to the 
initial passage of an absorbing cloud or streamline across the line of sight to the X-ray source. In the spectral analysis, this can be accounted for by an increase in the 
column density of the ionized absorber, by $\Delta N_{\rm H}\approx10^{24}$\,cm$^{-2}$, which as shown in Section~4 can account for the increase in 9 keV opacity and the enhanced variability in the F$_{\rm var}$ spectrum at this energy. 

From the timescale and magnitude of the absorber variability, it is possible to derive constraints on its sizescale, density and location. The thickness of a passing cloud or streamline can be estimated by $\Delta r = v_{\rm t} \Delta t$, where $v_{\rm t}$ is the transverse velocity of the gas cloud across the line of sight, which is initially assumed to be equal to the terminal wind speed ($v_{\rm t}=v_{\infty}$). Thus for $\Delta t=10$\,ks and $v_{\infty}=0.26c$, then $\Delta r = 8\times10^{13} \approx10^{14}$\,cm (or $\Delta r = 50R_{\rm g}$ for a $10^{7} {\rm M}_{\odot}$ black hole mass in \pg, e.g. \citealt{VestergaardPeterson06, Shen11}). The average hydrogen density of the gas is then $n=\Delta N_{\rm H}/\Delta r \approx 10^{10}$\,cm$^{-3}$. Given the density, then the radial distance can be estimated 
from the definition of the ionization parameter of $\xi = L_{\rm ion}/nR^2$, where $L_{\rm ion}$ is the 1--1000~Rydberg ionizing luminosity. Here $L_{\rm ion}=1.5\times10^{45}$\,erg\,s$^{-1}$ as determined from the broad--band fit in Section~5 to the 2022 slice A epoch\footnote{The 1--1000 Rydberg luminosity was calculated from extrapolating the broad--band power-law plus Comptonized disk model in the 2022 \xmm\ spectrum, towards 
lower energies. The extrapolated model is consistent with the photometry measured in the UV, from the simultaneous \xmm\ Optical Monitor data points.} 
Thus for the ionization parameter of $\log\xi=5$, as determined from the \textsc{xstar} fits in Section~4, the radial distance is estimated as $R\approx10^{15}$\,cm (or $\sim700R_{\rm g}$).  
Finally the characteristic length scale ($l$) of the absorber is given by the total duration ($\Delta t_{\rm d}$) of the absorption event, where $l= v_{\rm t} \Delta t_{\rm d}\sim5\times10^{14}$\,cm. Note this is a lower limit, as the minimum duration of the absorption dip is $t_{\rm d}=60$\,ks during slice B, as we do not see the recovery from the dip at the end of the 2022 observation.

Alternatively, we can relax the assumption of $v_{\rm t} = v_{\infty}$ and instead set it equal to the Keplerian rotation velocity at a radial distance $R$, where $v_{\rm K}=\sqrt{GM/R}$.  Substituting for $v_{\rm t}=v_{\rm K}$ into the above expressions (see \citealt{Risaliti05} and \citealt{ReevesLobbanPounds18}) then leads to a radial distance estimate of:-

\begin{equation}
R^{5/2} = (GM)^{1/2} \frac{L_{\rm ion} \Delta t}{\Delta N_{\rm H} \xi}.
\end{equation}

\noindent Thus for the above numbers, $R\approx5\times10^{14}$\,cm (or $\sim 300R_{\rm g}$), similar to the above estimate. At this distance and for $\log\xi=5$, $n=6\times10^{10}$\,cm$^{-3}$, 
while $v_{\rm K}\sim0.05c$ and $\Delta r = v_{\rm K} \Delta t \sim 2\times10^{13}$\,cm ($\sim15R_{\rm g}$). 

In either case above, the variability of the absorber could be explained by a transiting (or rotating) clump or streamline as part of an inhomogeneous wind. Given the above conditions, the absorption likely occurs  at a typical distance of a few hundred gravitational radii from the black hole and consists of clumps of thickness of a few tens of gravitational radii. Observationally, the increase in opacity in \pg\ is very similar to what is seen in the ultra fast outflow in \mcg\ \citep{Braito18}. There, a drop in flux and a corresponding increase in obscuration was observed in a 2015 {\it NuSTAR} observation, with the dip lasting for just over a day. The drop could be accounted for by an increase in opacity by a fast ($\sim0.1c$) wind, via an increase in its column and decrease in its ionization. Similarly, the location of the absorber was derived to be a few hundred gravitational radii, likely within a clumpy disk wind. 

These typical scales are in agreement with the hydrodynamical wind simulations of \citet{Mizumoto21}, who 
demonstrate that the bulk of the UFO wind absorption occurs at radii much larger than the launch radius, once the gas density is high enough and the ionization drops to values of $\log\xi=5$ or below. These authors applied their model to the well known ultra fast outflow in PG\,1211+143 \citep{Pounds03} and are able to reproduce the Fe K absorption structure in this QSO. 
Interestingly, \citet{Waters22} also suggest that clumps can form in highly supersonic outflows, as a result of the gas becoming thermally unstable. This occurs once the gas is fully accelerated close to its terminal speed rather than at the wind launching point, as otherwise acceleration can prohibit clumps from forming. These authors also suggest that intrinsic flux variability can enhance the clump formation, which may be especially relevant to NLS1s like \pg, which are highly variable in X-rays and can exhibit strong ultra fast outflows \citep{Hagino16, Parker17, Kosec18, ReevesBraito19}. 

\section{Conclusions}

New 2022 {\it NuSTAR} and \xmm\ observations of the narrow-lined Seyfert 1 galaxy, \pg, have revealed a highly variable wind. These revealed a much faster wind (of $v/c=0.26$) compared to an earlier 2017 
observation (with $v/c=0.1$), which overall was in a lower flux state. This is evidenced by the increased blue-shift of the iron K absorption trough when comparing these observations, from 7.5\,keV in 2017 to up to 9\,keV in 2022. 
Short term variability of the wind is observed in 2022, with the dip in the last 60\,ks of the {\it NuSTAR} observation coinciding with an increase in opacity of the fast disk wind. 
This can potentially be explained by denser clumps or streamlines of gas intercepting our line of sight to the source, within an inhomogeneous wind and on scales of a few hundred gravitational radii. 

Several open questions remain as to the origin of the wind variability in \pg. One is the cause of the drastic change in wind velocity and the blueshift of the iron K absorption profile. Overall the 2017 \xmm\ campaign caught the 
QSO in a lower overall flux, which also coincided with a period where the X-ray flux went through several dips (by up to an order of magnitude), as is seen in the Swift monitoring analyzed by \citet{Laurenti21}.  One possibility is the wind velocity reacts to the decrease in ionizing flux, becoming less efficiently accelerated, either by radiation or by a reduction in the overall magnetic activity of the corona, while in 2022 the wind is faster when the AGN is in a brighter X-ray flux state. Here, the ratio of the wind momentum rate to radiation force increases by a factor of five comparing the 2017 and 2022 (slice B) epochs, but does not appear to exceed unity overall and thus could be consistent with an increase in radiation driving. 

The second open question is the frequency and cause of the X-ray flux dips and whether they are connected to wind obscuration events, which can increase the X-ray opacity -- especially if the wind streamlines become Compton thick ($N_{\rm H}>10^{24}$\,cm$^{-2}$) to electron scattering and fully cover the X-ray source.  Future observations of \pg, via an approved $3\times80$\,ks joint \xmm\ and {\it NuSTAR} program in 2023 and coupled with daily {\it Swift} monitoring, will shed more light on how the wind reacts to the continuum.  In particular, the observations may provide insight into the frequency of dipping events, how or if the wind opacity changes as a result and whether there is any connection between the overall X-ray luminosity and the resultant wind velocity. 

\section{Acknowledgements}
We would like to thank Stuart Sim for the use of his disk wind radiative transfer code used in this paper. 
JR and VB acknowledge financial support through NASA grants 80NSSC22K0474 and 80NSSC22K0003. D.P acknowledges financial support from the CNES French space agency. 
Based on observations with {\it XMM-Newton}, an ESA science mission with instruments and contributions directly funded by ESA Member States and NASA and from the {\it NuSTAR} mission, a project led by the California Institute of Technology, managed by the Jet Propulsion Laboratory, and funded by NASA.
This research makes use of the NuSTAR Data Analysis Software (NuSTARDAS) jointly developed by the ASI Science Data Center and the California Institute of Technology.


\begin{thebibliography}

\bibitem[Boller et al.(1996)]{Boller96} Boller, T., Brandt, W.~N., \& Fink, H.\ 1996, \aap, 305, 53
\bibitem[Braito et al.(2018)]{Braito18} Braito, V., Reeves, J.~N., Matzeu, G.~A., et al.\ 2018, \mnras, 479, 3592 
\bibitem[Braito et al.(2021)]{Braito21} Braito, V., Reeves, J.~N., Severgnini, P., et al.\ 2021, \mnras, 500, 291
\bibitem[Braito et al.(2022)]{Braito22} Braito, V., Reeves, J.~N., Matzeu, G., et al.\ 2022, \apj, 926, 219.  
\bibitem[Chartas et al.(2002)]{Chartas02}Chartas, G., Brandt, W.~N., Gallagher, S.~C., \& Garmire, G.~P.\ 2002, \apj, 579, 169
\bibitem[Chartas \& Canas(2018)]{ChartasCanas18} Chartas, G. \& Canas, M.~H.\ 2018, \apj, 867, 103
\bibitem[Crenshaw et al.(2003)]{Crenshaw03} Crenshaw, D.~M., Kraemer, S.~B., \& George, I.~M.\ 2003, \araa, 41, 117
\bibitem[den Herder et al.(2001)]{denHerder01} den Herder, J.~W., Brinkman, A.~C., Kahn, S.~M., et al.\ 2001, \aap, 365, L7
\bibitem[Di Matteo et al.(2005)]{DiMatteo05} Di Matteo, T., Springel, V., \& Hernquist, L.\ 2005, \nat, 433, 604
\bibitem[Fabian(1999)]{Fabian99} Fabian, A. C., 1999, MNRAS, 308, L39
\bibitem[Ferrarese \& Merritt(2000)]{FerrareseMerritt00}Ferrarese L., Merritt D., 2000, ApJ, 539, 9
\bibitem[Fukumura et al.(2018)]{Fukumura18} Fukumura, K., Kazanas, D., Shrader, C., et al.\ 2018, \apjl, 864, L27
\bibitem[Garc{\'\i}a et al.(2014)]{Garcia14} Garc{\'\i}a, J., Dauser, T., Lohfink, A., et al.\ 2014, \apj, 782, 76
\bibitem[Gebhardt(2000)]{Gebhardt00}Gebhardt K., 2000, ApJ, 539, 13
\bibitem[Gofford et al.(2013)]{Gofford13}Gofford J., Reeves J. N., Tombesi F., et al., 2013, MNRAS, 430, 60
\bibitem[Gofford et al.(2014)]{Gofford14} Gofford, J., Reeves, J.~N., Braito, V., et al.\ 2014, \apj, 784, 77 
\bibitem[Grevesse \& Sauval(1998)]{GrevesseSauval98}Grevesse N., Sauval A. J., 1998, SSRv, 85, 161
\bibitem[Grupe et al.(2004)]{Grupe04} Grupe, D., Wills, B.~J., Leighly, K.~M., et al.\ 2004, \aj, 127, 156.
\bibitem[Hagino et al.(2016)]{Hagino16} Hagino, K., Odaka, H., Done, C., et al.\ 2016, \mnras, 461, 3954
\bibitem[Harrison et al.(2013)]{Harrison13} Harrison, F.~A., Craig, W.~W., Christensen, F.~E., et al.\ 2013, \apj, 770, 103 
\bibitem[Hopkins \& Elvis(2010)]{HopkinsElvis10}Hopkins P. F., Elvis M., 2010, MNRAS, 401, 7
\bibitem[Igo et al.(2020)]{Igo20} Igo, Z., Parker, M.~L., Matzeu, G.~A., et al.\ 2020, \mnras, 493, 1088.
\bibitem[Kaastra et al.(2000)]{Kaastra00} Kaastra, J.~S., Mewe, R., Liedahl, D.~A., et al.\ 2000, \aap, 354, L83
\bibitem[Kalberla et al.(2005)]{Kalberla05} Kalberla, P.~M.~W., Burton, W.~B., Hartmann, D., et al.\ 2005, \aap, 440, 775
\bibitem[Kallman et al.(2004)]{Kallman04} Kallman, T. R., Palmeri, P., Bautista, M. A., Mendoza, C., 
\& Krolik, J. H., 2004, ApJS, 155, 675
\bibitem[Kaspi et al.(2002)]{Kaspi02} Kaspi, S., Brandt, W.~N., George, I.~M., et al.\ 2002, \apj, 574, 643
\bibitem[King(2003)]{King03} King, A. R., 2003, ApJ, 596, L27
\bibitem[Kosec et al.(2018)]{Kosec18} Kosec, P., Buisson, D.~J.~K., Parker, M.~L., et al.\ 2018, \mnras, 481, 947 
\bibitem[Kosec et al.(2020)]{Kosec20} Kosec, P., Zoghbi, A., Walton, D.~J., et al.\ 2020, \mnras, 495, 4769.
\bibitem[Laurenti et al.(2021)]{Laurenti21} Laurenti, M., Luminari, A., Tombesi, F., et al.\ 2021, \aap, 645, A118.
\bibitem[Longinotti et al.(2015)]{Longinotti15} Longinotti, A.~L., Krongold, Y., Guainazzi, M., et al.\ 2015, \apjl, 813, L39
\bibitem[Luminari et al.(2018)]{Luminari18} Luminari, A., Piconcelli, E., Tombesi, F., et al.\ 2018, \aap, 619, A149
\bibitem[Matzeu et al.(2016)]{Matzeu16} Matzeu, G.~A., Reeves, J.~N., Nardini, E., et al.\ 2016, \mnras, 458, 1311 
\bibitem[Matzeu et al.(2017)]{Matzeu17} Matzeu, G.~A., Reeves, J.~N., Braito, V., et al.\ 2017, \mnras, 472, L15 
\bibitem[Matzeu et al.(2022)]{Matzeu22} Matzeu, G.~A., Lieu, M., Costa, M.~T., et al.\ 2022, \mnras, 515, 6172 
\bibitem[Mizumoto et al.(2021)]{Mizumoto21} Mizumoto, M., Nomura, M., Done, C., et al.\ 2021, \mnras, 503, 1442
\bibitem[Nardini et al.(2015)]{Nardini15} Nardini, E., Reeves, J.~N., Gofford, J., et al.\ 2015, Science, 347, 860
\bibitem[Parker et al.(2017)]{Parker17} Parker, M.~L., Alston, W.~N., Buisson, D.~J.~K., et al.\ 2017, \mnras, 469, 1553 
\bibitem[Parker et al.(2020)]{Parker20} Parker, M.~L., Alston, W.~N., Igo, Z., et al.\ 2020, \mnras, 492, 1363.
\bibitem[Pinto et al.(2018)]{Pinto18} Pinto, C., Alston, W., Parker, M.~L., et al.\ 2018, \mnras, 476, 1021 
\bibitem[Porquet et al.(2021)]{Porquet21} Porquet, D., Reeves, J.~N., Grosso, N., et al.\ 2021, \aap, 654, A89.
\bibitem[Porquet et al.(2018)]{Porquet18} Porquet, D., Reeves, J.~N., Matt, G., et al.\ 2018, \aap, 609, A42.
\bibitem[Pounds et al.(2003)]{Pounds03} Pounds, K.~A., Reeves, J.~N., King, A.~R., et al.\ 2003, \mnras, 345, 705 
\bibitem[Rakshit, Stalin \& Kotilainen(2020)]{Rakshit20} Rakshit, S., Stalin, C.~S., \& Kotilainen, J.\ 2020, \apjs, 249, 17.
\bibitem[Reeves et al.(2003)]{Reeves03} Reeves, J.~N., O'Brien, P.~T., \& Ward, M.~J.\ 2003, \apjl, 593, L65
\bibitem[Reeves et al.(2014)]{Reeves14} Reeves, J.~N., Braito, V., Gofford, J., et al.\ 2014, \apj, 780, 45 
\bibitem[Reeves, Lobban \& Pounds(2018)]{ReevesLobbanPounds18} Reeves, J.~N., Lobban, A., \& Pounds, K.~A.\ 2018, \apj, 854, 28
\bibitem[Reeves et al.(2018)]{Reeves18} Reeves, J.~N., Braito, V., Nardini, E., et al.\ 2018, \apj, 867, 38.
\bibitem[Reeves \& Braito(2019)]{ReevesBraito19} Reeves, J.~N. \& Braito, V.\ 2019, \apj, 884, 80
\bibitem[Reeves et al.(2021)]{Reeves21} Reeves, J.~N., Braito, V., Porquet, D., et al.\ 2021, \mnras, 500, 1974
\bibitem[Risaliti et al.(2005)]{Risaliti05} Risaliti, G., Elvis, M., Fabbiano, G., et al.\ 2005, \apjl, 623, L93
\bibitem[Saez \& Chartas(2011)]{SaezChartas11} Saez, C., \& Chartas, G.\ 2011, \apj, 737, 91
\bibitem[Schmidt \& Green(1983)]{SchmidtGreen83} Schmidt, M., \& Green, R.~F.\ 1983, \apj, 269, 352 
\bibitem[Shen et al.(2011)]{Shen11} Shen, Y., Richards, G.~T., Strauss, M.~A., et al.\ 2011, \apjs, 194, 45
\bibitem[Sidoli et al.(2001)]{Sidoli01} Sidoli, L., Oosterbroek, T., Parmar, A.~N., et al.\ 2001, \aap, 379, 540. 
\bibitem[Silk \& Rees(1998)]{SilkRees98} Silk, J., \& Rees, M. J., 1998, A\&A, 331, L1
\bibitem[Sim et al.(2008)]{Sim08}Sim, S. A., Long, K.S., Miller, L., \& Turner, T.J., 2008, MNRAS, 388, 611
\bibitem[Sim et al.(2010a)]{Sim10a} Sim, S.~A., Miller, L., Long, K.~S., Turner, T.~J., \& Reeves, J.~N.\ 2010, \mnras, 404, 1369 
\bibitem[Sim et al.(2010b)]{Sim10b}Sim S. A., Proga D., Miller L., Long K. S., Turner T. J., 2010, MNRAS, 408, 1396
\bibitem[Str{\"u}der et al.(2001)]{Struder01} Str{\"u}der, L., Briel, U., Dennerl, K., et al.\ 2001, \aap, 365, L18.
\bibitem[Titarchuk(1994)]{Titarchuk94} Titarchuk, L.\ 1994, \apj, 434, 570
\bibitem[Tombesi et al.(2010)]{Tombesi10} Tombesi, F., Cappi, M., Reeves, J.~N., et al.\ 2010, \aap, 521, A57 
\bibitem[Tremaine et al.(2002)]{Tremaine02} Tremaine, S., Gebhardt, K., Bender, R., et al.\ 2002, \apj, 574, 740 
\bibitem[Turner et al.(2001)]{Turner01} Turner, M.~J.~L., Abbey, A., Arnaud, M., et al.\ 2001, \aap, 365, L27
\bibitem[Vestergaard \& Peterson(2006)]{VestergaardPeterson06} Vestergaard, M., \& Peterson, B.~M.\ 2006, \apj, 641, 689 
\bibitem[Vaughan et al.(2003)]{Vaughan03} Vaughan, S., Edelson, R., Warwick, R.~S., et al.\ 2003, \mnras, 345, 1271 
\bibitem[Waters et al.(2022)]{Waters22} Waters, T., Proga, D., Dannen, R., et al.\ 2022, \apj, 931, 134
\bibitem[Wilms et al.(2000)]{Wilms00} Wilms, J., Allen, A., \& McCray, R. 2000, ApJ, 542, 914

\end{thebibliography}
\end{document}